\documentclass{article}

\usepackage{arxiv}

\usepackage[utf8]{inputenc} % allow utf-8 input
\usepackage[T1]{fontenc}    % use 8-bit T1 fonts
\usepackage{subfigure}
\usepackage{graphicx}
\usepackage{color}
\usepackage[section]{placeins}
\usepackage[normalem]{ulem}
\usepackage{amsmath,amsbsy,amssymb}
\usepackage{hyperref}

\title{The Kinematics of Lagrangian Flow Separation in External Aerodynamics}
\author{
Bjoern F. Klose \\
Department of Aerospace Engineering\\
San Diego State University\\
San Diego\\
\AND
Mattia Serra\\
School of Engineering and Applied Sciences\\
Harvard University\\
Cambridge\\
\AND
Gustaaf B. Jacobs\\
Department of Aerospace Engineering\\
San Diego State University\\
San Diego\\
}

\begin{document}
\maketitle

\begin{abstract}
Kinematic aspects of flow separation in external aerodynamics are investigated in the Lagrangian frame. Specifically, the initial motion of upwelling fluid material from the wall is related to the long-term attracting manifolds in the flow field. While the short-time kinematics are governed by the formation of a material spike upstream of the zero-skin-friction point and ejection of particles in direction of the asymptotic separation line, the trajectories of the fluid tracers are guided by attracting ridges in the finite-time Lyapunov exponents once they leave the vicinity of the wall.  The wall signature of this initial fluid upwelling event, the so-called \textit{spiking point} [Serra, M., Vetel, J., Haller, G., ``Exact theory of material spike formation in flow separation'', \textit{J. Fluid Mech.}, Vol. 845, 2018], is computed from the curvature of advected material lines and, for the first time, from high-order numerical derivatives of the wall-normal velocity obtained from direct numerical simulations of a circular cylinder and a cambered NACA 65(1)-412 airfoil.  As the spline-based boundary parametrization of the airfoil profile induces oscillations, the principle spiking point can be recovered robustly through appropriate filtering.  The short-term kinematics correlate strongly with the scaling lengths in the boundary layer.
\end{abstract}

\keywords{flow separation, Lagrangian, backbone of separation}

\section{Introduction} \label{introduction}
As practical small-scale flying devices proliferate and as interest in turbomachinery at
various scales develops, it becomes increasingly important to understand and characterize flows at moderate
Reynolds number.  In this flow regime, laminar 
boundary layer separation, reattachment, and transition results in significant
changes of the lift and drag, affecting the performance of the airfoil. The
control and prevention of flow separation can therefore yield a substantial
extension of the operating range of the device. The dynamics of flow separation
are highly non-linear, however,  making the design of an  effective passive and active flow
control system challenging.

While some flow control concepts work by completely removing or re-energizing
the separated fluid through suction and blowing (see e.g. Schlichting
\cite{schlichting}), other techniques take advantage of instabilities and
non-linearities in the flow by using more compact zero-net mass flux (ZNMF)
devices such as synthetic jets \cite{amitay98,amitay01, glezer02}. Given the
limited power of these ZNMF instruments, the design of an effective and
efficient control strategy should not only focus on directly changing the
global events in the flow, but rather on understanding and controlling 
the more subtle and unsteady features of separation.  \bigskip

In  steady flows,  separation  from a no-slip wall is 
well-known to be exactly identified by Prandtl's condition through
a point of zero skin friction and a negative friction gradient in
wall-tangential direction. For unsteady flows, 
similar first-principle criteria were only recently developed by Haller \cite{haller04}.
He proved that for time periodic flows, an objective
Lagrangian separation point is located at the averaged zero-skin-friction
location. Haller further showed that flow separation from a no-slip boundary
starts with an  upwelling of Lagrangian fluid tracers upstream of the separation
point and that those particles are drawn towards an unstable manifold in the
flow while they are ejected from the wall. This so-called asymptotic separation profile is
anchored at the separation point and it guides fluid particles 
as they break away in the vicinity of
the wall (see Haller \cite{haller04} and Weldon \textit{et al.}
\cite{WPJHH08}). 

\begin{figure}[htp]
	\subfigure{\includegraphics[width=0.45\textwidth]{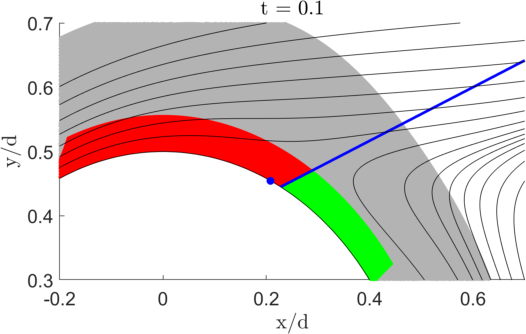}}
	\hfill
	\subfigure{\includegraphics[width=0.45\textwidth]{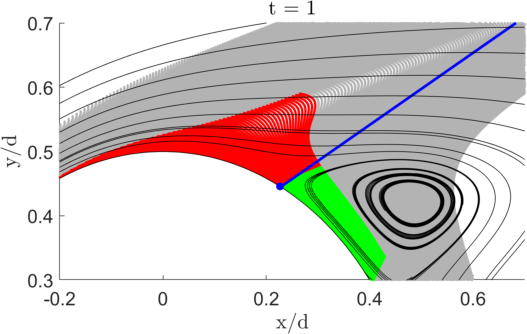}}
	\vskip\baselineskip
	\subfigure{\includegraphics[width=0.45\textwidth]{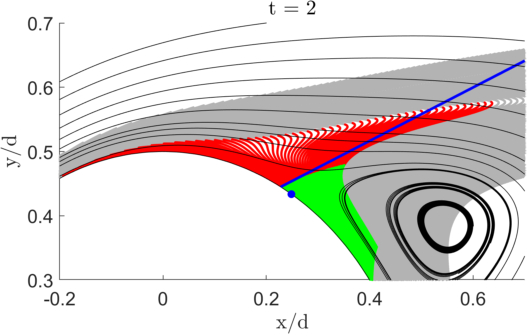}}
	\hfill
	\subfigure{\includegraphics[width=0.45\textwidth]{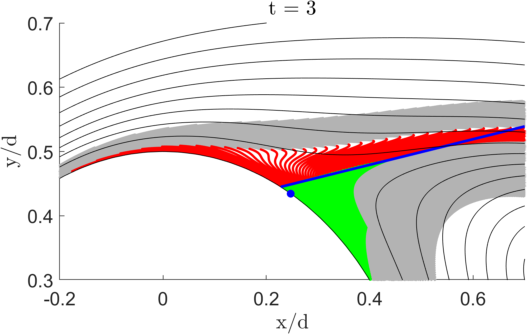}}
	\caption{Advection of particles over a cylinder near the separation point. Particles divided by linear separation line (blue) in upstream (red) and downstream (green). Zero-skin-friction point in blue, streamlines in black.}
	\label{fig:sepline}
\end{figure}

To illustrate this Lagrangian separation behavior, we consider the
time periodic flow over a circular cylinder in Figure \ref{fig:sepline}.
A set of fluid particle tracers is initialized in a layer parallel to the cylinder 
wall and is color-coded based on the linear approximation
of the dividing asymptotic separation line. 
%and advected with the time
%periodic flow. 
%The color-coded up- and downstream particles are separated by the line in the vicinity of the no-slip wall where the linear approximation is valid. 
As the particles are advected, they undergo
an upwelling motion, which is visible through an increasingly sharp spike in the 
material lines that are initially parallel to the wall.  
The spikes of particles are asymptotically drawn towards the  attracting  separation line. 
Mathematically, these attracting lines are interpreted as unstable manifolds.

In general, stable and unstable manifolds can be identified by extracting ridges
in the Finite-Time Lyapunov Exponent (FTLE) fields. These FTLE fields 
are determined from the maximum deformations in flow maps 
that are in turn constructed from integrated particle tracer fields
in forward and backward time, respectively. 
The ridges usually demarcate a hyperbolic Lagrangian Coherent Structure (LCS)
\cite{haller01,haller00}.  Although a hyperbolic Lagrangian Coherent Structure (LCS) can be identified
% GBJ: shadden reference
through local maxima in the FTLE field, Haller \cite{Haller02} shows that the FTLE field
has ridges in regions of high shear which are non-hyperbolic \cite{haller2011}. 

\begin{figure}[htp]
	\centering
		\includegraphics[width=0.70\textwidth]{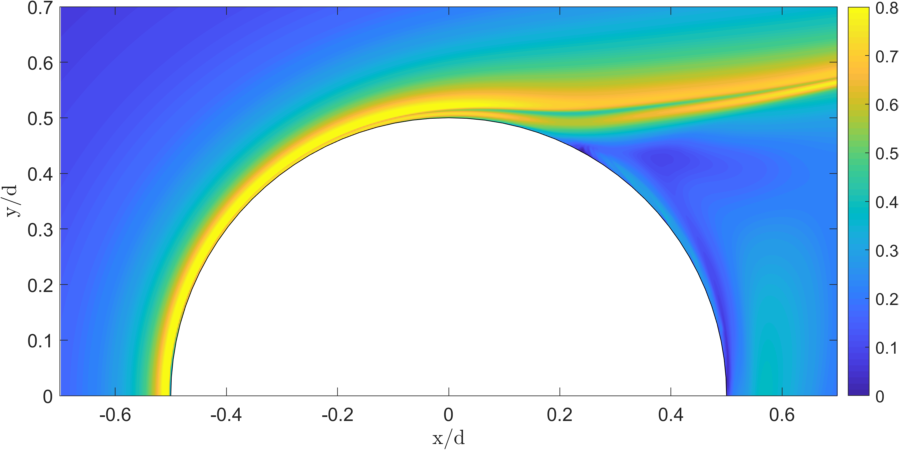}
	\caption{Backward-time FTLE around the upper half of a cylinder.}
	\label{fig:ftle}
\end{figure}

Even though hyperbolic LCS are mostly near zero-flux material lines \cite{shadden05},
they fall short in the identification of the
start of flow separation at the boundary wall. Because of the zero-velocity no-slip condition,
the wall is naturally a set of \textit{non}-hyperbolic fixed points.
As a result, the backward time (attracting) FTLE cannot intersect the
wall, but envelopes the aerodynamic body, as we show for the circular 
cylinder flow in Figure \ref{fig:ftle}.
The FTLE can, hence, only identify long-term attracting and repelling surfaces away
from the wall rather than the onset of separation.  

Based on the initial upwelling and subsequent spike formation described above,
Serra \textit{et al.} \cite{serra18} developed a kinematic methodology to
extract the \textit{Lagrangian backbone of separation}, i.e. the theoretical
centerpiece of the forming spike in wall-bounded flows from the analysis of the
curvature of material lines. The initial position of the backbone of separation
is a wall-transverse ridge of the Lagrangian curvature change field (see
below), while later positions can be captured by materially advecting the
initial position with the flow map. If the backbone connects to the wall, we
call the separation on wall, and the intersection point the Lagrangian spiking
point. This analysis therefore yields a criterion for determining the start of
flow separation in the Lagrangian frame.  Serra \textit{et al.} \cite{serra18}
directly relates the Lagrangian spiking point to higher-order derivatives of
the normal velocity at the wall, and thus provides a criterion for the origin
of flow separation in the kinematic sense using only wall-based quantities.
In the instantaneous limit, the Lagrangian backbone of separation turns into the Eulerian backbone of separation (Serra \textit{et al.} \cite{serra18}).
The work on material spike formation is elaborated by Serra \textit{et al.} \cite{serra19} for several example flows, including a separation bubble on a flat plate and a rotating cylinder. 
\bigskip

In this paper, we present a comprehensive kinematic study 
of Lagrangian flow separation in external aerodynamics by 
connecting  FTLE dynamics, the asymptotic separation line
and spike formation.
Using direct numerical simulations of a circular cylinder flow and the flow over a cambered NACA 65(1)-412 airfoil,
we show that  while the motion of fluid particles in the vicinity of the
no-slip wall is governed by the spike formation theory (Serra \textit{et al.}
\cite{serra18}) over short times, and over long times by the asymptotic
separation profile (Haller \cite{haller04}), the off-wall kinematics are
governed by long-term attracting LCSs in the flow field that can be extracted
from ridges in the backward-time FTLE. 
We find that the shape of the Lagrangian backbone of separation attains strong
bends along the boundary layer heights identified through momentum and
displacement thickness. These boundary layers thickness approximations are
based on kinetics arguments and typically involve a threshold value. The purely
kinematic Lagrangian backbone of separation, in contrast, is threshold free and
consistently distinguishes  on- and off-wall regions characterized by different
dynamics.

\begin{figure}[htp]
	\centering
		\includegraphics[width=0.70\textwidth]{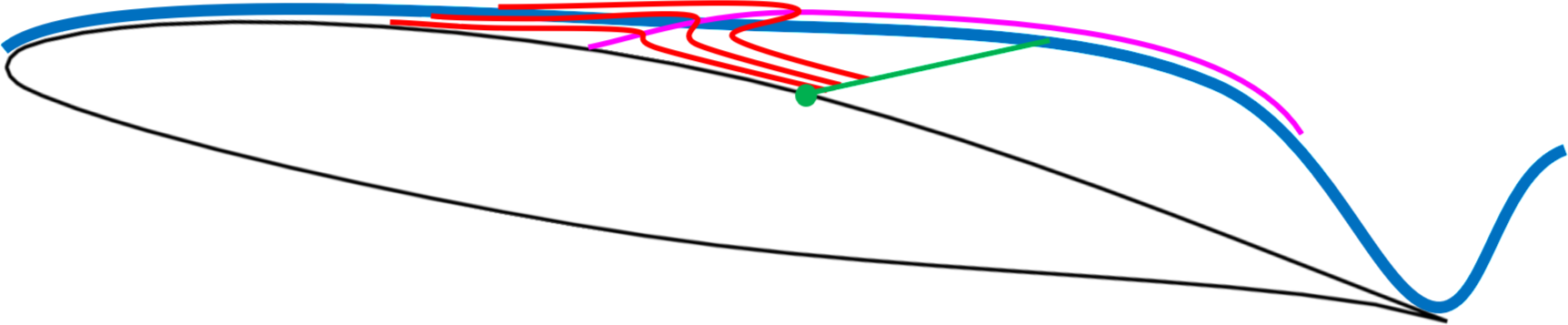}
	\caption{Schematic of the manifolds involved in Lagrangian flow separation: backward-time FTLE (blue), upwelling material lines (red) and associated Lagrangian backbone (magenta), linear separation profile and asymptotic separation point (green). Schematic is not to scale.}
	\label{fig:schematic}
\end{figure}

A schematic of the principal mechanisms of separation in steady or periodic
flows with an asymptotic mean is presented in Figure \ref{fig:schematic}. The
attracting LCS, which is identified from the backward-time FTLE ridge (blue),
attracts the upwelling fluid material (red) from the wall.   This ridge 
does not intersect with the wall, but rather develops along the separated shear layer. The Lagrangian backbone of
separation (magenta) is the theoretical centerpiece of the spike and intersects
the boundary at the so-called spiking point. Attracted by the hyperbolic LCS,
the backbone then aligns with the FTLE ridge once the Lagrangian fluid tracers
have left the vicinity of the wall. Downstream of the spiking point, the
asymptotic separation profile (green) is anchored at the location of averaged
zero skin friction and oriented in the direction of particle break-away.

We further identify the wall signature of upwelling material lines, i.e. the spiking
point \cite{serra18}, from Lagrangian quantities and, for the first time, from
high-order wall-based velocity derivatives. 
By matching the spiking points extracted from the Lagrangian curvature change of
material lines and Eulerian on-wall velocity derivatives of the circular
cylinder, we verify the criterion proposed by Serra \textit{et al.}
\cite{serra18}. The cylinder case benefits from an analytically known
boundary with constant curvature. To extend the test to a non-analytic wall
representations, the flow around the airfoil with a cubic-spline
based boundary is analyzed. Although the material lines show the formation of a single
spike at mid-cord characterized by a severe curvature change, very weak
curvature ridges occur upstream of the asymptotic location of separation. The
higher-order on-wall derivatives reflect these ridges and show their relation
to the piece-wise linear curvature of the wall. Applying a filter with a kernel
based on the distance between supporting points of the spline, we recover the
spiking point of the principal separation event from the on-wall velocity
derivatives and match the wall intersection of the Lagrangian backbone of
separation. The robustness of this method to local oscillations or noise is an
important result that will benefit future applications, given that many
engineering applications rely on surface representations through splines.

We further show that singular points, such as the stagnation point at the
leading edge, must be excluded from the analysis of material spike formation,
as they can induce fluid upwelling without separation.  Barring this
limitation, the material spike formation provides wall-based and short-term
information that remains hidden in the backward-time FTLE and the asymptotic
separation profile. The new information about the material spike
adds a valuable piece to the picture of Lagrangian flow separation and is a
promising tool in the design of flow controllers. 

%As a last point in the present study, we re-evaluate the relation between the material spike formation and the FTLE. While Serra \textit{et al.} found no direct connection, our results show that even though ridges in curvature and FTLE indeed deviate for short integration intervals, the spiking motion moves towards the FTLE ridge as they gradually align with increasing integration time. The on-wall conection of the curvature ridge, however, remains hidden in the FTLE field. 
\bigskip

The governing equations and the numerical model are given in Section \ref{methodology}. In Section \ref{setup} we outline the setup of our computations and the results are discussed in the subsequent part. A summary and conclusion is given in Section \ref{conclusion}.

\section{Methodology} \label{methodology}
\subsection*{Governing Equations}
We consider the compressible Navier-Stokes equations for conservation of mass, momentum and energy, which can be written in non-dimensional form as a system of equations where the flux vector is divided into an advective (superscript \textit{a}) and a viscous part (superscript \textit{v}):

\begin{equation}
\frac{\partial\mathbf{Q}}{\partial t} + \mathbf{F}_x^a + \mathbf{G}_y^a - \frac{1}{Re_f}\left(\mathbf{F}_x^v + \mathbf{G}_y^v\right) = 0.
	\label{eq:NS}
\end{equation}
The solution and flux vectors are
\begin{equation}
\mathbf{Q} = \left[\,\rho \quad \rho u \quad \rho v \quad \rho E\,\right]^T,
\end{equation}
\begin{equation}
\mathbf{F}^a = \left[\,
\rho u \quad p {+} \rho u^2 \quad \rho u v \quad u(\rho E {+} p)
\,\right]^T,
\end{equation}
\begin{equation}
\mathbf{G}^a = \left[\,
\rho v \quad \rho v u \quad p {+} \rho v^2 \quad v(\rho E {+} p)
\,\right]^T,
\end{equation}
\begin{equation}
\mathbf{F}^v = \left[\,
0 \quad \tau_{xx} \quad \tau_{yx} \quad u \tau_{xx} {+} v \tau_{yx} {+} \frac{\kappa}{\left(\gamma -1\right)Pr M_f^2} T_x
\,\right]^T,
\end{equation}
\begin{equation}
\mathbf{G}^v = \left[\,
0 \quad \tau_{xy} \quad \tau_{yy} \quad u \tau_{xy} {+} v \tau_{yy} {+} \frac{\kappa}{\left(\gamma -1\right)Pr M_f^2} T_y
\,\right]^T,
\end{equation}

and the equation of state follows as
\begin{equation}
p = \frac{\rho T}{\gamma M_f^2}.
\end{equation}

All quantities are non-dimensionalized with respect to a characteristic, problem-dependent length
scale, reference velocity, density, and temperature yielding the non-dimensional Reynolds number,
$Re_f$ and Mach number, $M_f$.

\subsection*{Discontinuous Galerkin Spectral Element Method}
On the computational domain, the system of equations \ref{eq:NS} is spatially approximated using a discontinuous Galerkin spectral element method (DGSEM) and integrated in time with a 4th-order explicit Runge-Kutta scheme. Gauss-Lobatto quadrature nodes are used for the spatial integration and a kinetic-energy conserving split-form approximation of the advective volume fluxes ensures stability of the scheme through cancellation of aliasing errors from the the non-linear terms.
For a detailed description of the scheme, we refer to Kopriva \cite{kopriva}, Gassner \textit{et al.} \cite{gassner16} and Klose \textit{et al.} \cite{kloseAIAA19}.
%A kinetic energy conserving split-form approximation for the inviscid fluxes is used in order to ensure numerical stability of the scheme \cite{gassner16}.

%%%%%%%%%%%%%%%%%%%%%%%%%%%%%%%%%%%%%%%%%%%%%%%%%%%%%%%%%%%%%%%%%%%%%%%%%%%%%%%

%\subsection*{Lagrangian Coherent Structures}
%%\subsubsection*{Velocity-Gradient Based Methods}
%%A range of different methods has been used to extract the topology of a given flow field. A popular choice is to identify coherent structures as regions of constant vorticity $\mathbf{\omega} = \nabla \times \mathbf{u}$. However, for shear flows such as wall-bounded flows, high velocity gradients do not exclusively emerge from the swirling motion of the fluid. 
%%
%%The Q-criterion, defined as the second invariant of the velocity gradient, is another common method to extract vortical structures from the flow field. 
%%\begin{equation}
%%Q = \frac{1}{2}\left((\nabla\cdot \mathbf{u})^2 + |\mathbf{\Omega}|^2 - |\mathbf{S}|^2\right),
%%\end{equation}
%%For $Q > 0$, only regions where the vorticity $\mathbf{\omega} = 2 \mathbf{\Omega}$ exceeds the strain rate $\mathbf{S}$ are taken into account, hence revealing structures that might be obscured by the iso-vorticity method.
%
\subsection*{Finite-Time Lyapunov Exponent}
We extract structures and patterns from flow field data using a Finite-Time Lyapunov Exponent contour field \cite{shadden05}.
The FTLE, which characterizes the maximal stretching of infinitesimal fluid volumes over a given time interval, is determined by tracing fluid particles over time and subsequently computing the deformation tensor induced by the flow map. 

We express the particle trajectories as
\begin{equation}
\mathbf{x}\left(\mathbf{x}_0,t_0;T\right) = \mathbf{x}_0 + \int_{t_0}^{t_0+T} \mathbf{v}\left(\mathbf{x}\left(\tau;\mathbf{x}_0,t_0\right),\tau\right)d\tau,
\end{equation}
from which the flow map $\mathbf{F}$ is defined:
\begin{equation}
\mathbf{F}_{t_0}^{t}\left(\mathbf{x_0},t_0;T\right) \equiv \mathbf{x}\left(\mathbf{x}_0,t_0;T\right).
\end{equation}
From the deformation gradient tensor $\nabla\mathbf{F}_{t_0}^{t}$, the right Cauchy-Green strain tensor $\mathbf{C}_{t_0}^{t}=[\nabla\mathbf{F}_{t_0}^{t}]^*\nabla\mathbf{F}_{t_0}^{t}$ can be used to compute the strain in the Lagrangian frame. With the largest eigenvalue of the strain tensor $\lambda_2\left(\mathbf{C}_{t_0}^{t}(\mathbf{x}_0)\right)$, the finite-time Lyapunov exponent field is defined as
\begin{equation}
\Lambda_{t_0}^{t}(\mathbf{x}_0)=\frac{1}{\left|t-t_0\right|}\ln \sqrt{\lambda_2(\mathbf{x}_0)}.
\end{equation}
This  FTLE identifies  the highest Lagrangian rate of stretching in the flow field. Tracing fluid particles
forward or backward in time, ridges of the FTLE field can be used to identify
hyperbolic repelling and attracting Lagrangian coherent structures (see Haller
\cite{haller01, Haller02}, and Nelson and Jacobs \cite{nelson15} for a more
detailed description).

Although ridges in the backward-time FTLE are associated with separation (see 
 and Lipinski \textit{et al.} \cite{lipinski}), the no-slip condition makes the wall a set of non-hyperbolic fixed points, inhibiting any transverse intersections with FTLE ridges.
Furthermore, the birth of a material spike is not a stretching-dominated phenomenon, but rather the outcome of an interplay of stretching and rotation objectively captured by the material curvature (Serra \textit{et al.} \cite{serra18}). The strain-based FTLE field is therefore not suited to detect the start of Lagrangian flow separation on no-slip boundaries. 
%
%Lipinski \textit{et al.} \cite{lipinski} further showed that material is ejected from the wall in the presence of an unstable manifold. So, it can be inferred that flow separation is related to ridges in the backward-time FTLE, while the forward-time FTLE shows flow reattachment.

%
\subsection*{Separation Point and Angle}
%In the Lagrangian frame, boundary layer separation is defined as the ejection of fluid particles located in the vicinity of a no-slip wall. These particles are drawn towards an attracting material line in the flow field, referred to as separation line. Given the fact that attractors in the flow field can be highlighted by ridges in the backward-time FTLE (whereas ridges in the forward-time FTLE are linked to repellers), the backward-time FTLE is in a powerful tool to identify the separation line. However, given the zero-velocity condition at the wall, no material can be attracted or repelled, so that the LCS never intersects with the wall and thus cannot be used to identify the separation point. 

Haller \cite{haller04} shows that for flows with an asymptotic mean, such as periodic flows, the asymptotic separation point $\gamma$ is located at the integrated zero-skin-friction point
\begin{equation}
\frac{1}{t_1-t_0}\int_{t_0}^{t_1} c_f\left(\gamma,t\right)dt = 0.
	\label{eq:sep_point}
\end{equation}
Haller further derives an analytic expression for the separation profile which is a wall-bounded unsteady manifold along which fluid particles are ejected from the wall into the free-stream. The slope, or separation angle, of this line can be computed just by evaluating integrated values of the pressure and skin friction data at the wall:
\begin{equation} \label{eq:alpha}
\tan\left(\alpha\left(t_0\right)\right) = -\lim_{T \to -\infty}\frac{3\int_{t_0}^T\tau_x\left(\gamma,t\right)dt}{\int_{t_0}^T\left[p_x\left(\gamma,y_w,t\right)+3\tau_x\left(\gamma,t\right)\int_{t_0}^t\frac{1}{\mu}\tau\left(\gamma,s\right)ds\right]dt}.
\end{equation}
Here, the \textit{x} coordinate refers to wall tangential direction and \textit{y} points in wall-normal direction. The separation angle $\alpha$ is the angle to the tangent of the wall at the separation point.
Using both, the separation point and the separation angle, a linear approximation of the separation profile can be constructed. 

%%%%%%%%%%%%%%%%%%%%%%%%%%%%%%%%%%%%%%%%%%%%%%%%%%%%%%%%%%%%%%%%%%%%%%%%%%%%%%%
%\subsection*{Free Stream Turbulence}
%Following a method introduced by Kanchi \textit{et al.} \cite{kanchi13}, a time correlated signal can be generated at the inlet boundary condition by computing the autocorrelation function $R\left(\tau\right)$ if the integral time scale $T_i$ of the free stream turbulence is known. The integral time scale can be computed from a velocity fluctuation data as
%\begin{equation}
%T_i = \int_{0}^{\infty} R\left(\tau\right)d\tau = \int_{0}^{\infty} \frac{\overline{u\left(t\right)'u\left(t+\tau\right)'}}{\overline{u\left(t\right)'^2}} d\tau
%\end{equation}

\subsection*{The Lagrangian Backbone of Separation and the Spiking Point}
Flow separation is invariably characterized by the ejection of fluid particles from a no-slip wall. While the long-time (asymptotic) behavior of these particles are governed by attracting LCSs  in the flow field, the onset of separation is not related to asymptotic structures. Serra \textit{et al.} \cite{serra18} show that the formation of a material spike is characterized by high folding induced by the flow on material lines close to the wall (Figure \ref{fig:schematic}), which appears at a different location - generally upstream - compared to the asymptotic separation point (e.g., the zero-skin-friction point in the case of steady flows). This deformed spike, then, eventually converges to the breakaway from the wall along the corresponding long-term separation structure. The materially evolving set of points forming the centerpiece of the separation spike (magenta curve in Figure \ref{fig:schematic}) is also referred to as the \textit{backbone of separation} \cite{serra18}. 

Following Serra \textit{et al.} \cite{serra18}, separation is on-wall if the backbone has a transverse intersection with the non-slip boundary, off-wall otherwise. 
We note that such a distinction is not postulated a priori based on heuristic arguments, but rather is an outcome of the theory proposed  in \cite{serra18}. Using a coordinate system $[s, \eta]$ in direction tangential and normal to the wall respectively, we compute the Lagrangian curvature change relative to the initial curvature $\bar{\kappa}_{{t_0}}^{t_0+T} := \kappa^{t_0+T}_{t_0} - \kappa_0$ in a neighborhood of the no-slip boundary foliated by a set of material lines initially parallel to the wall, parametrized by $\mathbf{r}_\eta(s),\ s \in[s_1, s_2] \subset \mathbb{R}, \ \eta \in[0, \eta_{1}] \subset \mathbb{R}$.
Such a foliation enslaves the initial local tangent $\mathbf{r}'_\eta$ and curvature $\kappa_{{0}_{\eta}}$ to the position $\mathbf{r}_\eta$, making therefore $\bar{\kappa}_{{t_0}}^{t_0+T}$ a function of $t_{0},T$ and of the initial configuration $\mathbf{r}_\eta$ only. Here $(\cdot)':=\tfrac{d}{ds}(\cdot)$. The $\bar{\kappa}_{t_0}^{t_0+T}$ field can be directly computed from the flow map $\mathbf{F}_{t_0}^{t_0+T}$ using the relation
\begin{multline} \label{eq:kappa_hat}
\bar{\kappa}_{t_0}^{t_0+T} = \frac{\left\langle\left(\nabla^2\mathbf{F}_{t_0}^{t_0+T}(\mathbf{r}_\eta)\mathbf{r}_\eta'\right)\mathbf{r}_\eta', \mathbf{R}\nabla\mathbf{F}_{t_0}^{t_0+T}(\mathbf{r}_\eta)\mathbf{r}_\eta'\right\rangle}{\left\langle\mathbf{r}_\eta', \mathbf{C}_{t_0}^{t_0+T}(\mathbf{r}_\eta)\mathbf{r}_\eta'\right\rangle^{3/2}} -
\kappa_{0_\eta}\left[\frac{det\left(\nabla\mathbf{F}_{t_0}^{t_0+T}(\mathbf{r}_\eta)\right)\left\langle\mathbf{r}_\eta',\mathbf{r}_\eta'\right\rangle^{3/2}}{\left\langle\mathbf{r}_\eta', \mathbf{C}_{t_0}^{t_0+T}(\mathbf{r}_\eta)\mathbf{r}_\eta'\right\rangle^{3/2}} - 1\right],
\end{multline}
where $\langle \cdot , \cdot \rangle$ denotes the inner product; $(\nabla^2 \mathbf{F}_{t_0}^{t}(\mathbf{r}_\eta)\mathbf{r}'_\eta)_{ij}=\sum\limits_{k=1}^{2}{\partial_ {jk}F_{t_{0}}^t}_{i}(\mathbf{r}_\eta)r'_{{\eta}_k},\ i,j\in\{1,2\}$, and $\mathbf{R}$ is the rotation matrix defined as
\begin{align}
\mathbf{R} := \begin{bmatrix}
0 & 1 \\
-1 & 0
\end{bmatrix}.
\end{align}\label{kt}
We note that with a clockwise parametrization of the no-slip boundary, $\mathbf{R}\mathbf{r}'_\eta$ is the vector normal to the initial material line, pointing towards the boundary. 
The initial position $\mathcal{B}(t_0)$ of the Lagrangian backbone of separation -- i.e., the theoretical centerpiece of the material spike over $[t_0,t_0+T]$ -- is then defined as a positive-valued wall-transverse ridge of the $\bar{\kappa}_{t_0}^{t_0+T}$ field (Serra \textit{et al.} \cite{serra18} for details).  
%In other words, $\mathcal{B}(t_0)$ is the set of points where $\bar{\kappa}_{t_0}^{t_0+T}$ will attain a local maximum with respect to wall-parallel direction, i.e. the set of points $s\in [s_1,s_2]$, $\eta\in [0,\eta_1]$:
%\begin{equation}
%\label{eq:cond_lagr}
%\begin{aligned}
%\mathcal{B}(t_0):=
%\begin{cases}
%\partial_s \overline{\kappa}_{t_{0}}^{t_{0}+T}(s,\eta)=0,\ \ \ \ \eta\in (0,\eta_1]\\
%\partial_{ss} \overline{\kappa}_{t_{0}}^{t_{0}+T}(s,\eta)<0,\ \ \ \eta\in (0,\eta_1]\\
%\overline{\kappa}_{t_{0}}^{t_{0}+T}(s,\eta)>0,\ \ \ \ \ \ \ \eta\in (0,\eta_1]\\
%(s_p,\eta),\ \ \ \ \ \ \ \ \ \ \ \ \ \ \ \ \  \ \eta=0.\\
%\end{cases}
%\end{aligned}
%\end{equation}
Later position of the backbone $\mathcal{B}(t)$ can be computed by materially advecting $\mathcal{B}(t_0)$, i.e., letting $\mathcal{B}(t) := \mathbf{F}_{t_0}^{t}(\mathcal{B}(t_0)),\ t\in [t_0,t_0+T]$.
If $\mathcal{B}(t_0)$ connects to the wall transversally, the intersection point is called the Lagrangian spiking point and is defined by
\begin{equation}
(s_p,0):=\mathcal{B}(t_0)\cap \text{no-slip wall}.
\label{eq:SpikPointDef}
\end{equation}

% If $\mathcal{B}(t_0)$ connects to the wall, the separation is on-wall, otherwise off-wall. In the case of on-wall separation, the intersection of $\mathcal{B}(t_0)$ and the wall defines the \textit{Lagrangian spiking point} $s_p$, which captures the on-wall signature of the spike formation (Fig. \ref{fig:illustr}). Analytic expression for $s_p$ are available in \cite{Serra2018}.

Serra \textit{et al.} \cite{serra18} derived also alternative exact formulas for the Lagrangian spiking point using only on-wall Eulerian quantities in the case of steady, time periodic and generally aperiodic flows (cf. Table \ref{tab:sepPointLagrFormula}). 
\begin{table}
	\centering
	\begin{tabular}{c|c}
		\multicolumn{2}{c}{$\mathbf{Lagrangian\ spiking\ point}:(s_p,0)$}
		\tabularnewline
		\hline 
		$\nabla\cdot \mathbf{v}\neq 0$  &  $\nabla\cdot \mathbf{v}=0$ \\
		\hline 	 
		$\begin{cases}
\int_{t_0}^{t_0+T}\partial_{sss\eta}\hat{v}\left(s_p,0,t\right)=0, \\
\int_{t_0}^{t_0+T}\partial_{ssss\eta}\hat{v}\left(s_p,0,t\right)>0, \\
\int_{t_0}^{t_0+T}\partial_{ss\eta}\hat{v}\left(s_p,0,t\right)<0.
		\end{cases}$ 
		& 	
		$\begin{cases}
\int_{t_0}^{t_0+T}\partial_{sss\eta\eta}\hat{v}\left(s_p,0,t\right)=0, \\
\int_{t_0}^{t_0+T}\partial_{ssss\eta\eta}\hat{v}\left(s_p,0,t\right)>0, \\
\int_{t_0}^{t_0+T}\partial_{ss\eta\eta}\hat{v}\left(s_p,0,t\right)<0,
		\end{cases}$\\
	\end{tabular}
	\caption{Equations determining the Lagrangian spiking point for generally aperiodic compressible (left) and incompressible (right) flows on a no-slip boundary in terms of on-wall Eulerian quantities. $\hat{v}$ indicates the velocity direction normal to the wall.}\label{tab:sepPointLagrFormula}
\end{table}
%Instead of using the curvature change $\bar{\kappa}_{t_0}^{t_0+T}$ from Lagrangian fluid tracers and the associated deformation gradient $\nabla\mathbf{F}$, the Eulerian spiking point is based on the curvature rate $\dot{\kappa}_{t_0}$ and the rate-of-strain tensor $\mathbf{S} = (\nabla\mathbf{u}+\nabla\mathbf{u}^T)/2$ (see \cite{serra18} for details).
Finally, in the instantaneous limit ($T=0$), the Lagrangian backbone of separation and spiking point
turns into their correspondent Eulerian versions (Serra \textit{et al.} \cite{serra18}).

\noindent In Table \ref{tab:sepPointLagrFormula}, $\hat{v}$ indicates the velocity in normal direction to the wall, and can be computed from the inner product:
\begin{equation} \label{eq:uhat}
\hat{v} = \langle \mathbf{v}, \mathbf{n}\rangle = u n_x + v n_y,\quad \mathbf{n} := \mathbf{R}\frac{\mathbf{r}'_\eta}{\vert \mathbf{r}'_\eta \vert},\quad \vert \mathbf{r}'_\eta \vert = \sqrt{\langle \mathbf{r}'_\eta,\mathbf{r}'_\eta \rangle}.
\end{equation}
Assuming a curved, parametrized boundary $\mathcal{W}(s)$, the normal vector $\mathbf{n}$ at each collocation point $\mathbf{x}_i$ is the vector pointing to the closest intersection point with the boundary $\mathcal{W}$ and found by minimizing the distance function $d_i(s) = \vert \mathbf{x}_i-\mathcal{W}(s)\vert$.
If we assume a continuous and sufficiently smooth function $\hat{v}$, we can switch the order of differentiation in Table \ref{tab:sepPointLagrFormula} and compute the normal gradients first. Once the normal velocity is obtained at each collocation point, the gradient can be computed by multiplication with the derivative matrix $\mathcal{D}$. Using the spectral operator $\mathcal{D}$ gives the derivatives in x- and y-direction in the polynomial order of the scheme. 
\begin{equation}
\nabla \hat{v} = \mathcal{D} \hat{v}.
\end{equation}

The directional derivative of the normal velocity $\hat{v}$ in direction of the wall-normal vector $\mathbf{n}$ is computed as
\begin{equation} \label{eq:gradn}
\partial_\eta\hat{v} = \nabla_\mathbf{n}\hat{v} = \nabla\hat{v}\cdot\mathbf{n} = \left(\partial_x \hat{v}\right) n_x + \left(\partial_y \hat{v}\right) n_y.
\end{equation}

This relation allows us to calculate the normal derivatives everywhere in the flow field from the velocity gradient and the normal vector at each point. Once $\partial_\eta\hat{v}$ or $\partial_{\eta\eta}\hat{v}$ are determined, the derivatives tangential to the wall can be computed subsequently.

\section{Problem Setup} \label{setup}
The canonical circular cylinder flow is computed at a Reynolds number of $Re_d = 100$ based on a cylinder diameter of unity and Mach number of 0.1, rendering compressibility effects negligible. The computational domain is divided into 347 quadrilateral elements and the solution is approximated with a 16\textsuperscript{th} order polynomial. This accounts to a total of 100,283 collocation points.
At the outer boundaries, a free stream condition is applied while the cylinder is approximated with curved element faces and an adiabatic no-slip wall.
402,201 Lagrangian particles are initialized in 201 wall-parallel lines around the cylinder with a spacing of $\Delta/d$ = 0.001 between each line.

\begin{figure}[htp]
	\centering
		\includegraphics[width=0.50\textwidth]{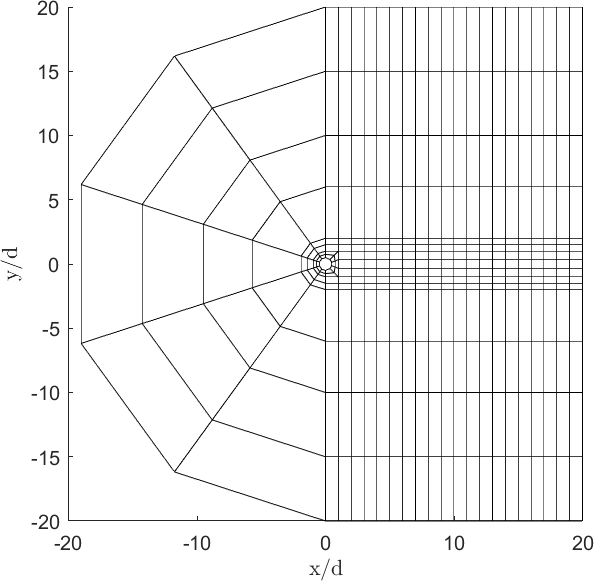}
	\caption{2D computational domain for circular cylinder.}
	\label{fig:cylinder}
\end{figure}

\begin{figure}[htp]
	\centering
		\includegraphics[width=0.60\textwidth]{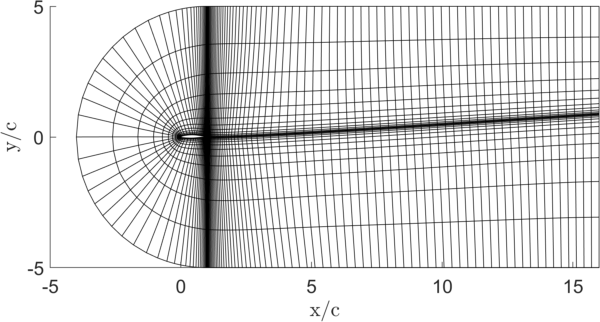}
	\caption{2D computational domain for NACA 65(1)-412. Only elements without interior Gauss-Lobatto nodes are shown.}
	\label{fig:grid}
\end{figure}

The flow over a NACA 65(1)-412 airfoil is simulated at a Reynolds number based on the chord length of $Re_c = 20,000$ and a Mach number of $M = 0.3$. The Mach number is relatively low ensuring a nominally incompressible flow, but it is high enough to prevent stability issues pertaining to the explicit time  integration we use. 
The computational domain is given in Figure \ref{fig:grid} and consists of 2,256 quadrilateral elements, with the dimensions of the domain being adopted from Jones \textit{et al.} \cite{jones08}. The boundary elements are curved and fitted to a spline representing the airfoil's surface according to Nelson \textit{et al.} \cite{nelson16}. Boundary conditions at outer edges of the computational domain are specified as free-stream boundaries while the airfoil surface is treated as a non-slip, adiabatic wall. The solution vector is approximated with a 16\textsuperscript{th} order polynomial, giving a total of 651,984 collocation points in the domain.
1,005,201 Lagrangian particles are initialized in 201 wall-parallel lines around the airfoil with a spacing of $\Delta/c$ = 0.0002 between each line. 

In both simulations, Lagrangian particles are tracked by spectrally interpolating the velocity field for each particle and numerically integrating $\dot{\mathbf{x}}\left(\mathbf{x},t\right) = \mathbf{v}\left(\mathbf{x}\left(t\right),t\right)$ using  a 3\textsuperscript{rd} order Adam-Bashfort scheme. The \textit{gslib} library is used for efficient particle tracking and velocity interpolation, as described by Mittal \textit{et al.} \cite{mittal18}.

The wall-normal derivatives in Table \ref{tab:sepPointLagrFormula} are computed within the DGSEM solver. 
With Equations \ref{eq:uhat} and \ref{eq:gradn}, the quantities $\hat{v}_\eta$ and $\hat{v}_{\eta\eta}$ can be spectrally computed in each element using the operators available in the DGSEM framework and subsequently interpolated to the wall. The derivatives in wall-tangential direction can either be computed within the DGSEM solver or as part of the post-processing work.
Given the sensitive nature to numerical noise of second and higher derivatives, a smoothing filter is applied to the DNS output data as a post-processing step.

\section{Results and Discussion} \label{results}
\subsection*{Cylinder Flow}
\begin{figure}[htp]
	\centering
	\includegraphics[width=0.80\textwidth]{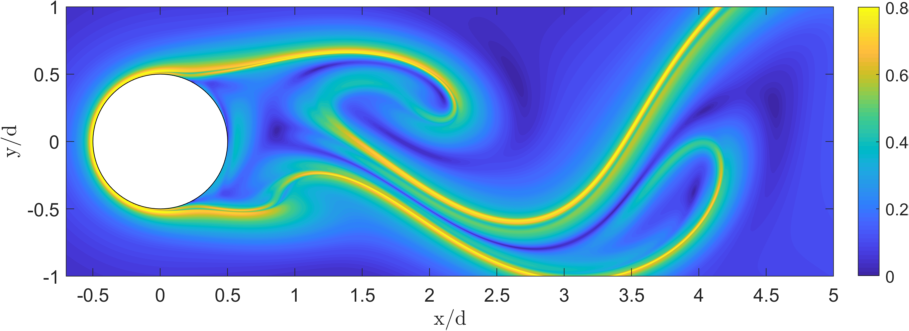}
	\caption{Backward-time FTLE from integration over one vortex shedding period.}
	\label{fig:cyl_flow}
\end{figure}

To study the kinematics of flow separation, we consider a cylinder flow at $Re_d = 100$. 
Ridges in the FTLE field show a flow pattern that is well-known to be dominated by a pair of counter-rotating vortices alternately shedding in a regular manner from the top and bottom of the cylinder with a period of
approximately six convective time units
\cite{Williamson96,RG19}. 
A snapshot of the backward-time FTLE (Figure \ref{fig:cyl_flow}) reveals the long-term attracting LCSs in the wake, which highlight the edges of the advected vortices.  
Although this LCS is  associated with separation (see Lipinski \textit{et al.} \cite{lipinski})
and with early vortex formation and shedding \cite{RG19}, 
the FTLE ridge cannot intersect with the cylinder wall
but rather envelopes the body. This is a direct consequence of the no-slip condition at the wall
and non-hyperbolicity, as was explained 
in the introduction. 
The exact on-wall origin of separation  can hence not be identified solely based on a  strain-based FTLE field.
A more rigorous analysis of the near-wall flow field is required. 
To this end, we first determine the asymptotic separation
point and line \cite{haller04} and then compute the Lagrangian curvature change $\bar{\kappa}_{t_0}^{t_0+T}$
and associated spiking dynamics.  Later, we  will relate the material spiking and the FTLE. 

The averaged zero-skin-friction point is determined according to Eq. \ref{eq:sep_point} with
the temporal mean of the skin friction coefficient over one vortex shedding period.
It is located at $x/d = 0.23$, 
 approximately half way between center and the rear end of the cylinder. 
 The angle of the separation line 
 with respect to the tangent of the cylinder surface at the
 separation point is determined with Eq. \ref{eq:alpha}.
 It  oscillates periodically  between 34$^\circ$ and 57$^\circ$.
 We use the angle and separation point to create a linear
 approximation of the unstable manifold to which fluid particles that eject from the wall
 are asymptotically attracted.

\begin{figure}[htp]
	\subfigure{\includegraphics[width=0.45\textwidth]{img/Cylinder/sepline_top0010.png}}
	\hfill
	\subfigure{\includegraphics[width=0.45\textwidth]{img/Cylinder/sepline_top0020.png}}
	\vskip\baselineskip
	\subfigure{\includegraphics[width=0.45\textwidth]{img/Cylinder/sepline_top0030.png}}
	\hfill
	\subfigure{\includegraphics[width=0.45\textwidth]{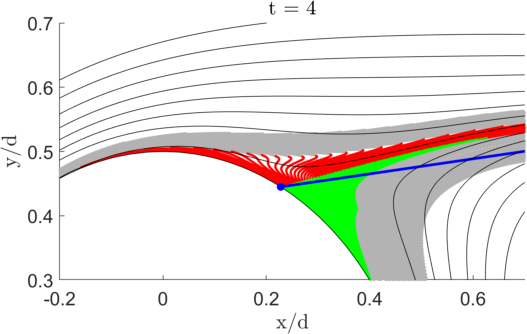}}
	\caption{Advection of particles over the cylinder near the separation point. Particles divided by linear separation line (blue) in upstream (red) and downstream (green). Zero-skin-friction point in blue, streamlines in black.}
	\label{fig:sepline_cyl}
\end{figure}

 The near-wall dynamics are visualized in Figure \ref{fig:sepline_cyl}, where
 color-coded fluid tracers, the linear separation profile and instantaneous
 streamlines are plotted for different integration times, $T$.
 Particles up- and downstream of the line
undergo an initial upwelling (spiking) and are drawn towards the unstable manifold.
 
\begin{figure}[htp]
	\centering
	\subfigure{\includegraphics[height=0.2\textheight]{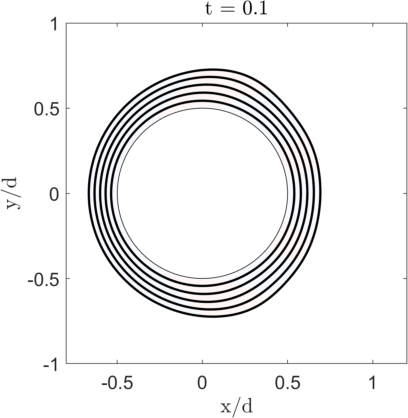}}
	\hspace{.5in}
	\subfigure{\includegraphics[height=0.15\textheight]{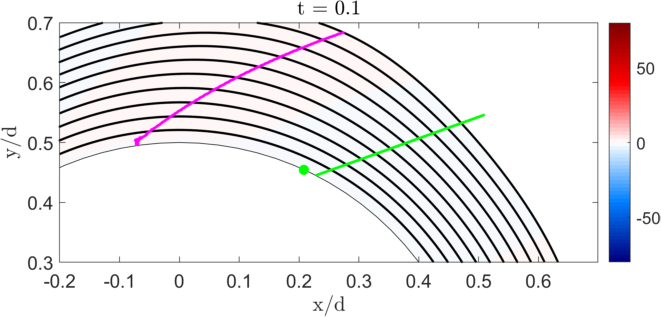}}
	\par
	\subfigure{\includegraphics[height=0.2\textheight]{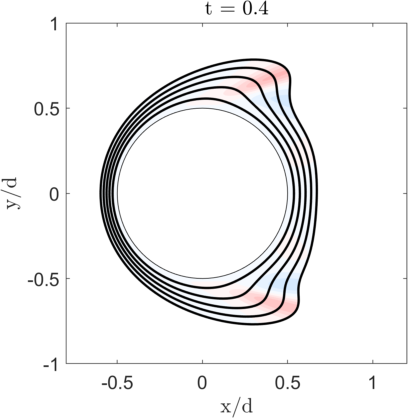}}
	\hspace{.5in}
	\subfigure{\includegraphics[height=0.15\textheight]{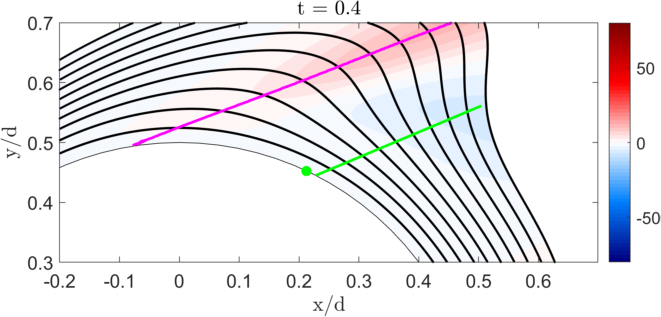}}
	\par
	\subfigure{\includegraphics[height=0.2\textheight]{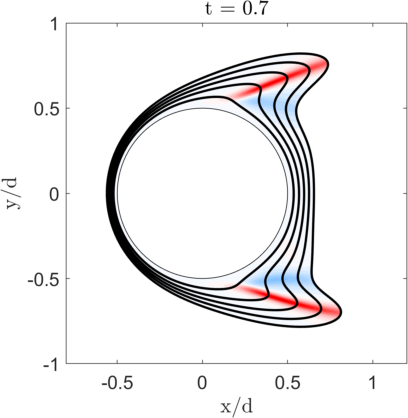}}
	\hspace{.5in}
	\subfigure{\includegraphics[height=0.15\textheight]{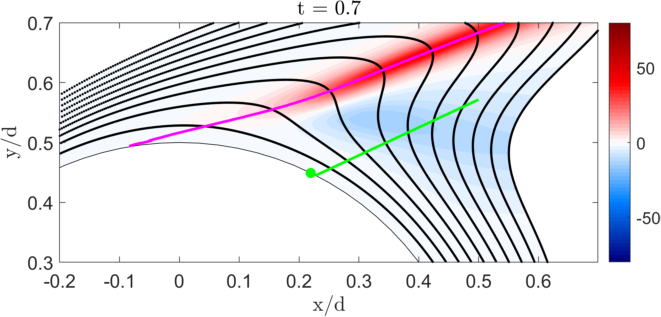}}
	\par
	\subfigure{\includegraphics[height=0.2\textheight]{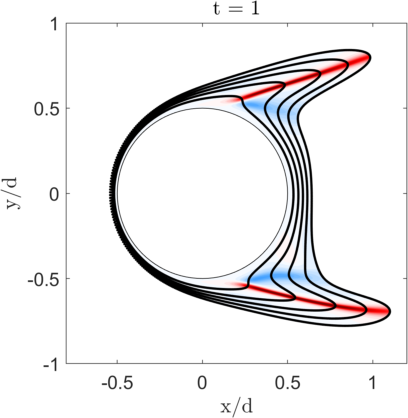}}
	\hspace{.5in}
	\subfigure{\includegraphics[height=0.15\textheight]{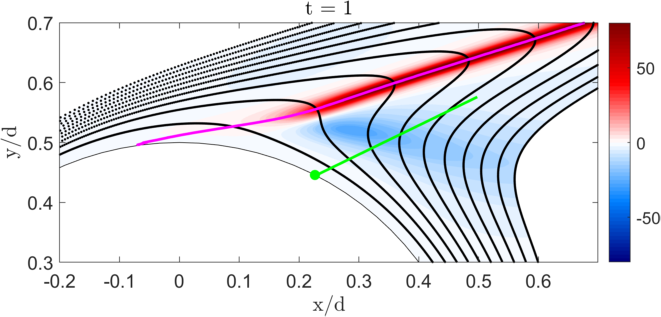}}
	\caption{Advection of material lines and the curvature field $\bar{\kappa}_{t_0}^{t},\ t=t_0+T$ around a cylinder for different integration times. The backbone $\mathcal{B}(t)$ is highlighted in magenta. Linear separation line and zero-skin-friction point in green.}
	\label{fig:matline_cyl}
\end{figure}

To identify the onset of flow separation, i.e. the origin of material spiking, 
we extract a backbone from the evolution of the material lines  through ridges
in the corresponding advected curvature field $\bar{\kappa}_{t_0}^{t_0+T}$. 
We plot the curvature field  for integration times of \textit{T} = [0.1, 0.4, 0.7, 1.0], and $t_0=0$ in Figure \ref{fig:matline_cyl}. Note that we first compute $\bar{\kappa}_{t_0}^{t_0+T}$, which is a scalar field based at $t_0$, and then advect it with $\mathbf{F}_{t_0}^{t_0+T}$. The latter operation reflects the material property of lines
and the backbone, $\mathcal{B}(t_0+T) := \mathbf{F}_{t_0}^{t_0+T}(\mathcal{B}(t_0))$. 
The backbone, $\mathcal{B}(t)$, shown in magenta. %in Figure \ref{fig:matline_cyl}.
For reference, we also plot the instantaneous zero-skin-friction point and the linear approximation of the separation profile in green. 

The evolution of material lines in Figure \ref{fig:matline_cyl} show that the backbone profile $\mathcal{B}(t)$ is correctly placed along the local spikes of material lines and intersects with the wall shortly upstream of the center of the cylinder. The separated fluid tracers then follow the direction of the linearly approximated separation profile.
While the asymptotic separation profile provides information about the long-term behavior of separating fluid tracers, the initial material spike formation remains hidden and can only be extracted from analysis of the curvature scalar field.

\begin{figure}[htp]
	\subfigure{\includegraphics[width=0.45\textwidth]{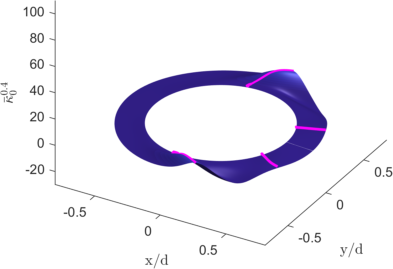}}
	\hfill
	\subfigure{\includegraphics[width=0.45\textwidth]{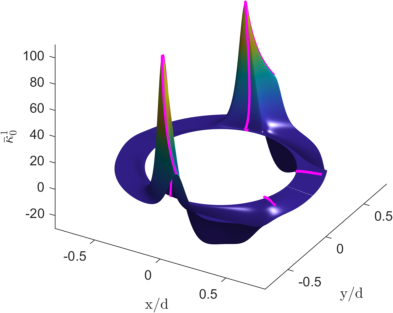}}
	\caption{Surface plot of the curvature scalar fields $\bar{\kappa}_{0}^{0.4}$ and $\bar{\kappa}_{0}^{1}$. Backbone of separation in magenta.}
	\label{fig:surf_cyl}
\end{figure}

The curvature change field $\bar{\kappa}_{t_0}^{t_0+T}$ for
integration time intervals of \textit{T} = 0.4 and 1.0 
in Figure \ref{fig:surf_cyl} reveal a total number of four Lagrangian backbones.
%of separation are extracted as positive wall-transverse ridges of the
%$\bar{\kappa}_{t_0}^{t_0+T}$ surfaces, 
Two originate from the top and the
bottom of the cylinder and evolve along a dominant, growing ridge in the
curvature field driven by the separation of the boundary layer. The two other
backbones are located within the recirculation region in the separated cylinder
wake. They are based on much weaker curvature ridges 
and we therefore deem them of secondary interest in the onset of separation.

\FloatBarrier

\subsubsection*{Spiking phenomenon and FTLE}
\begin{figure}[htp]
	\subfigure{\includegraphics[width=0.45\textwidth]{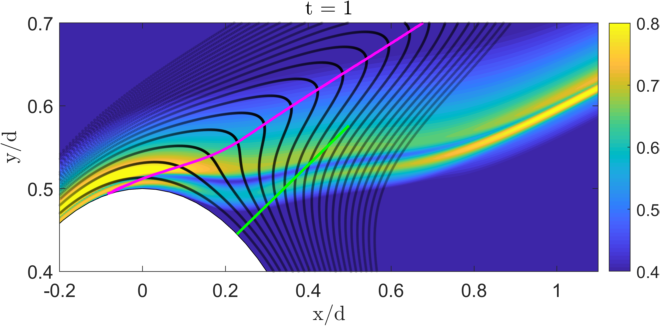}}
	\hfill
	\subfigure{\includegraphics[width=0.45\textwidth]{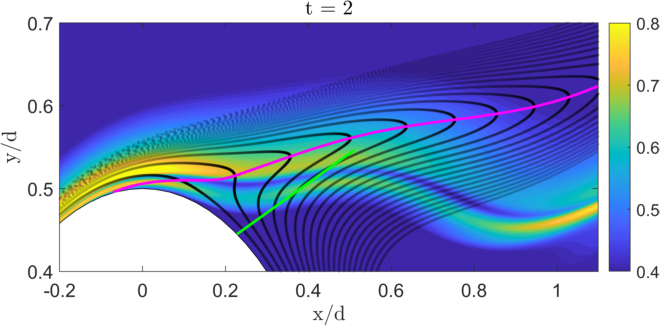}}
	\vskip\baselineskip
	\subfigure{\includegraphics[width=0.45\textwidth]{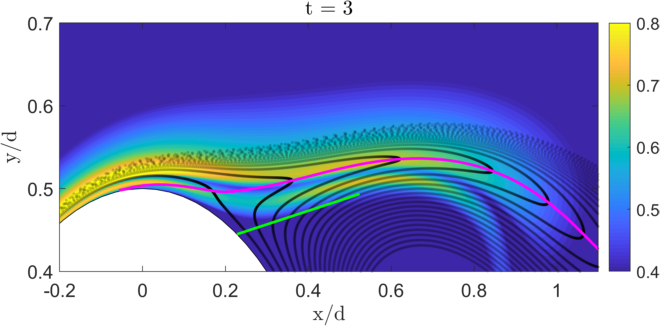}}
	\hfill
	\subfigure{\includegraphics[width=0.45\textwidth]{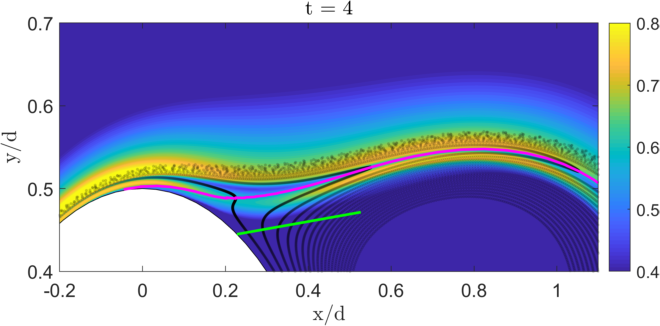}}
	\caption{Backward-time FTLE field (contour plot) computed from \textit{t} to \textit{t} $-$ \textit{T} over \textit{T} = 6. Advected material lines from 0 to \textit{t} in black and the Lagrangian backbone of separation in magenta. Asymptotic separation profile in green. Y-axis stretched.}
	\label{fig:ftle_cyl}
\end{figure}

The spike formation, which occurs over short time, is hidden to the FTLE field \cite{serra18}. 
For longer integration times, however, the material spike, governed by off-wall dynamics, converge to the attracting backward-time FTLE ridge (Fig.\ \ref{fig:schematic}).  
For the cylinder case, we visualize   the spatial relation between the backbone, material lines and the backward-time FTLE field at different time instances in Figure \ref{fig:ftle_cyl}.   
In this figure, a time interval of one vortex shedding period, \textit{T} = 6, is used to compute the FTLE field. 
Material lines and the backbone are advected from \textit{t} = 0 to \textit{t} = 1, 2, 3, and 4 (black). 

\begin{figure}[htp]
	\centering
	\includegraphics[width=0.70\textwidth]{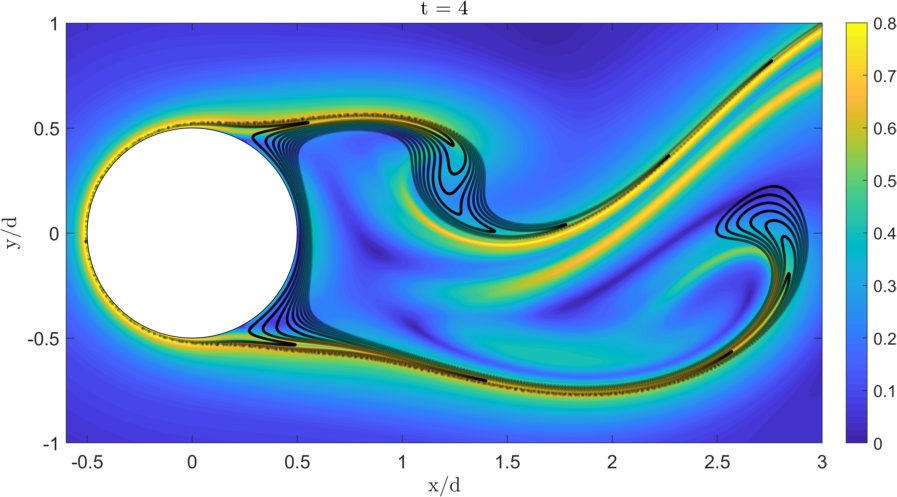}
	\caption{Backward-time FTLE and material lines in the wake.}
	\label{fig:ftle_wake_cyl}
\end{figure}

Initially, the fluid tracers undergo an upward motion in transverse direction to the cylinder and the backbone along the material spike crosses the FTLE ridge (\textit{t} = 1). As the integration time increases, however, the material lines bend downwards (\textit{t} = 2) and gradually align with the unstable manifold for \textit{t} $\geq$ 3. 
The long-term manifold identified through the backward-time FTLE ridge attracts the separating fluid material
and  gradually aligns with the material backbone.
The trace of the separated fluid in the wake follows the same pattern and shows long-term sharp spikes along dominant FTLE ridges, as illustrated in Figure \ref{fig:ftle_wake_cyl}.

%The response of the emerging particle tracers to the attracting LCS shows that FTLE ridges and the associated unstable manifolds in the flow field extract the off-wall dynamics of the material spike formation and represent fundamental structures in the analysis of Lagrangian flow separation. 
The above results highlight that the Lagrangian backbones of separation and the FTLE provide critical complementary structures in the analysis of Lagrangian flow separation.
While the initial motion through upwelling of fluid material can only be determined through the analysis of the curvature change field \cite{serra18}, the long-term off-wall dynamics are governed by the FTLE.
A combination of both methodologies therefore, together with the asymptotic separation line, gives a complete picture of the kinematics of separation (see Figure \ref{fig:schematic}).

\FloatBarrier

\subsubsection*{Extraction of Spiking Points}

The spiking points, $s_p$, 
are the wall signatures of material upwelling and can either be identified from the intersection of a wall-transverse curvature change ridge with the boundary (Eq. \eqref{eq:SpikPointDef}) or from on-wall Eulerian derivatives of the wall-normal velocity (Table \ref{tab:sepPointLagrFormula}).
Here, we extract $s_p$ using the criterion for incompressible flows, since
the flow  with a free-stream Mach number of \textit{M} = 0.1 is nearly incompressible 

\begin{figure}[htp]
\centering
	\subfigure[]{\includegraphics[width=0.35\textwidth]{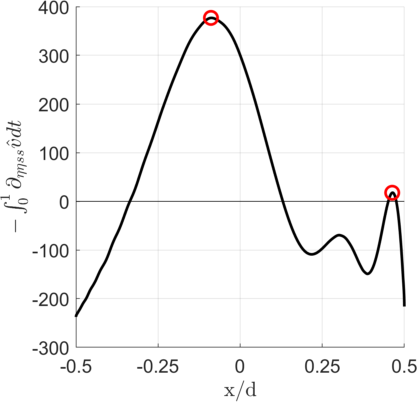}}
	\hspace{.5in}
	\subfigure[]{\includegraphics[width=0.4\textwidth]{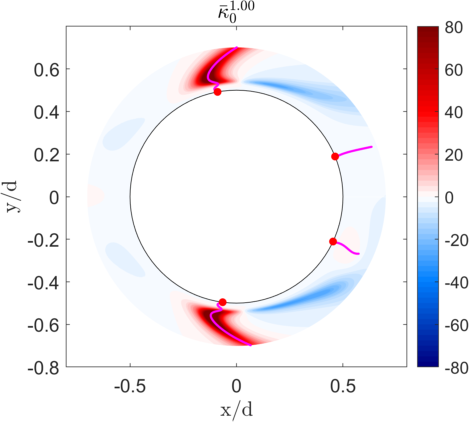}}
	\caption{(a) $-\int_{0}^{1}\partial_{\eta\eta ss}\hat{v} dt$ with spiking points in red. (b) Curvature change field $\bar{\kappa}_{0}^{1}$ with backbones (magenta) and Lagrangian spiking points (red) identified from Eulerian on-wall quantities.}
	\label{fig:vnn_cyl}
\end{figure}

From the condition specified in Table \ref{tab:sepPointLagrFormula}, the
spiking points are located at minima of the integrated derivatives of the
normal velocity in normal and tangential direction,
$\int_{t_0}^{t_0+T}\partial_{\eta\eta ss}\hat{v} dt$. We plot this function in
Figure \ref{fig:vnn_cyl} (a) for the upper half of the cylinder and an interval
of \textit{T} = 1. The resulting spiking points are indicated with red circles.
Figure \ref{fig:vnn_cyl} (b) shows the curvature change field
$\bar{\kappa}_{0}^{1}$ and the Lagrangian backbones of separation
$\mathcal{B}(t_0)$ in magenta. The spiking points identified from the normal
velocity derivatives are plotted as red dots at the boundary and match exactly
with the intersection of the backbones and the wall. Through the agreement of
the spiking points determined from Eulerian on-wall quantities and the
alternative Lagrangian definition (Eq. \eqref{eq:SpikPointDef}), here,
we verify the theory by Serra \textit{et al.} \cite{serra18}
\textit{for the first time}, i.e. we  confirm the theory that material upwelling in the Lagrangian frame can be captured also by using  wall-based Eulerian quantities only.  

\begin{figure}[htp]
	\subfigure{\includegraphics[width=0.45\textwidth]{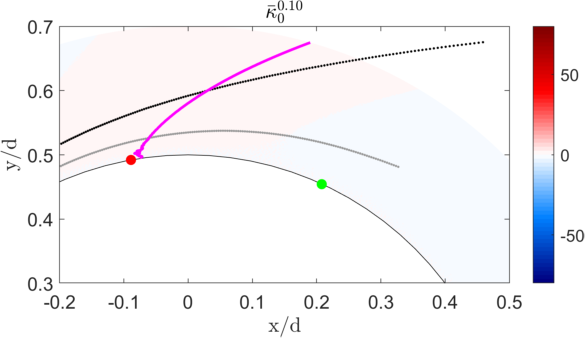}}
	\hfill
	\subfigure{\includegraphics[width=0.45\textwidth]{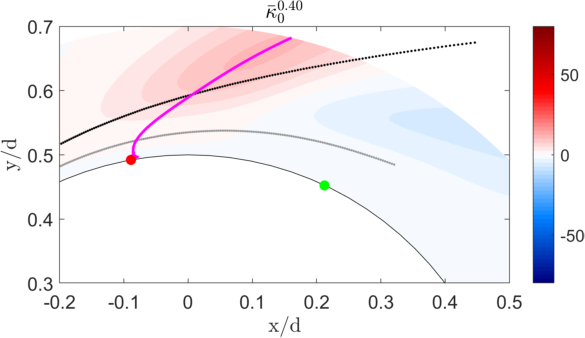}}
	\vskip\baselineskip
	\subfigure{\includegraphics[width=0.45\textwidth]{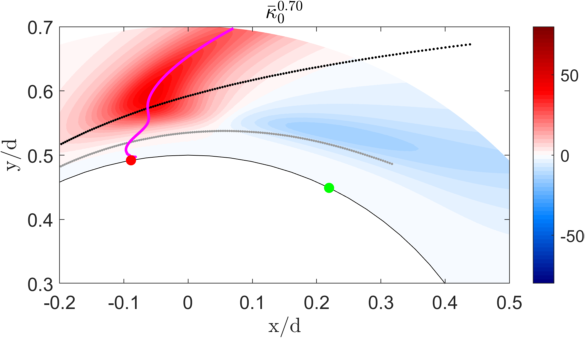}}
	\hfill
	\subfigure{\includegraphics[width=0.45\textwidth]{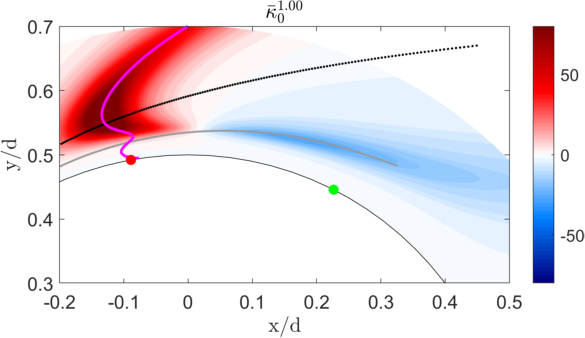}}
	\caption{Lagrangian curvature change field with the corresponding backbone of separation (magenta) and Lagrangian spiking points (red) identified from Eulerian on-wall quantities for different integration times. Zero-skin-friction point in green, boundary layer displacement thickness in black and momentum thickness in grey.}
	\label{fig:kappa_cyl}
\end{figure}

Figure \ref{fig:kappa_cyl} shows the curvature change field, based at the
initial time, on the upper side of the cylinder for increasing integration
times, together with the Lagrangian spiking points from on-wall quantities
(red), backbones of separation (magenta), and boundary layer approximations
based on the momentum and displacement thickness in grey and black
respectively. These plots are based on the same particle trace that is used in
the previous Figures \ref{fig:matline_cyl}, \ref{fig:surf_cyl}, and
\ref{fig:vnn_cyl} (b).  The Lagrangian spiking point $s_p$ is located at $x/d$
= -0.09, which places it far upstream of the asymptotic separation point ($x/d$
= 0.23) and, remarkably, even upstream from the cylinder center. 

We find that there is a strong correlation between the curvature change field and
the boundary layer scaling thicknesses, such as the displacement
thickness and momentum loss thickness \cite{schlichting}.
Figure \ref{fig:kappa_cyl} shows that, as the integration time
increases, ridges of $\bar{\kappa}_{t_0}^{t_0+T}$ form and develop a peak at
the intersection with the displacement thickness (black line). Within the
momentum thickness layer (grey line), the curvature of the ridge abruptly
decreases.  The dependence of the backbone of separation on the displacement
and momentum thickness is a remarkable result, as boundary layer thicknesses
follow kinetic arguments and typically involve thresholds parameters.
Inflection of the backbone of separation, in contrast, despite being threshold
free and purely kinematic, accurately separate  on- and off-wall regions
characterized by different dynamics. 
We are currently exploring this correlation in
material and plan to report on this in the near future.

We note that even though the curvature change ridge develops a `nose' and moves upstream with increasing integration time, the backbone $\mathcal{B}(t_0)$ maintains its original on-wall signature and intersects the wall at the spiking points identified by the criteria in Table \ref{tab:sepPointLagrFormula}. 

\FloatBarrier

\subsection*{Airfoil Flow}

For a more complex and encompassing external aerodynamics test case, we study the
kinematics of flow separation on a cambered NACA 65(1)-412 airfoil at a
chord-based Reynolds number of $Re_c = 20,000$ and 4$^\circ$ angle of attack.
The low Reynolds airfoil flow is characterized by boundary layer separation at mid-cord,
a recirculation region downstream of the separation location and a Von-Karmann-type vortex
shedding in the wake, resulting in a time-periodic flow pattern with a period
of \textit{T} = 0.36.  

\begin{figure}[htp]
	\centering
	\includegraphics[width=0.80\textwidth]{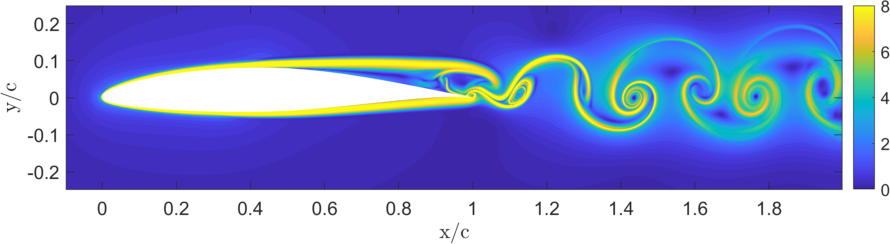}
	\caption{Backward-time FTLE from integration over one vortex shedding period.}
	\label{fig:airfoil_flow}
\end{figure}

A snapshot of the backward-time FTLE (Figure
\ref{fig:airfoil_flow}) visualizes
the separated shear layer and the edges of the shedded and advected vortices.

\begin{figure}[htp]
	\subfigure{\includegraphics[width=0.45\textwidth]{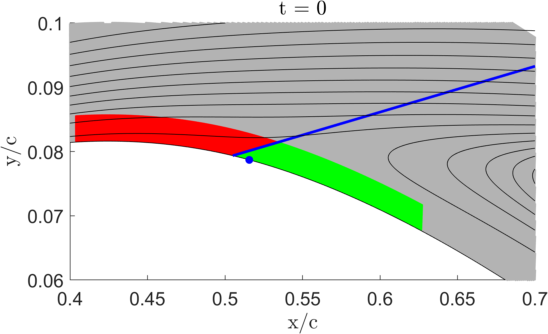}}
	\hfill
	\subfigure{\includegraphics[width=0.45\textwidth]{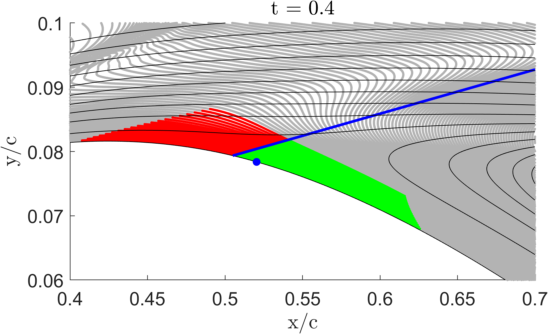}}
	\vskip\baselineskip
	\subfigure{\includegraphics[width=0.45\textwidth]{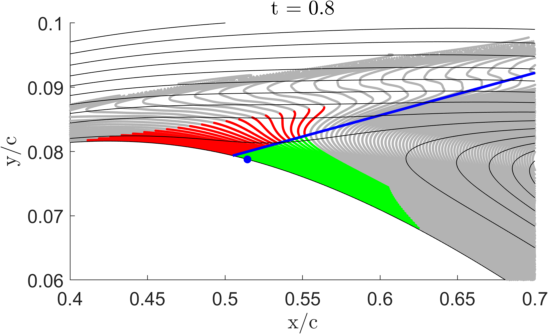}}
	\hfill
	\subfigure{\includegraphics[width=0.45\textwidth]{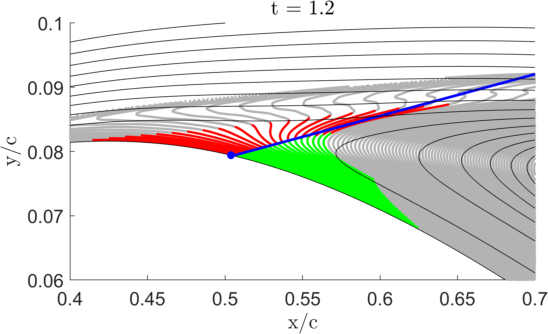}}
	\caption{Advection of particles over the airfoil near the asymptotic separation point. Particles divided by the asymptotic linear separation line (blue) in upstream (red) and downstream (green) locations. Zero-skin-friction point in blue, streamlines in black.}
	\label{fig:sepline_airf}
\end{figure}

The asymptotic separation point is computing using the mean over one vortex shedding period,
and is located at the averaged zero-skin-friction point at $x/c$ = 0.50, i.e.  exactly at mid-cord, slighly 
behind the maximum thickness location of the airfoil ($x/c$ = 0.4). This is in accordance with 
thre result reported in Nelson \textit{et al.} \cite{nelson16} and Kamphuis \textit{et al.} \cite{kamphuis}.
The angle of the separation line with respect to the tangent of the airfoil
surface periodically oscillates between 7.05$^\circ$ and 7.5$^\circ$.
%and provides a linear approximation of the wall-based unstable manifold that ejects
%fluid particles along the separation profile. 
These near-wall dynamics are summarized in Figure \ref{fig:sepline_airf}, where color-coded fluid tracers,
the asymptotic separation profile and instantaneous streamlines are plotted for
different integration times. 
Similar to the cylinder flow, the particles
upstream of the asymptotic separation point undergo an upwelling motion and
form a sharp spike that will be later guided by an attracting LCS in the flow.

\begin{figure}[htp]
	\centering
	\subfigure{\includegraphics[width=0.8\textwidth]{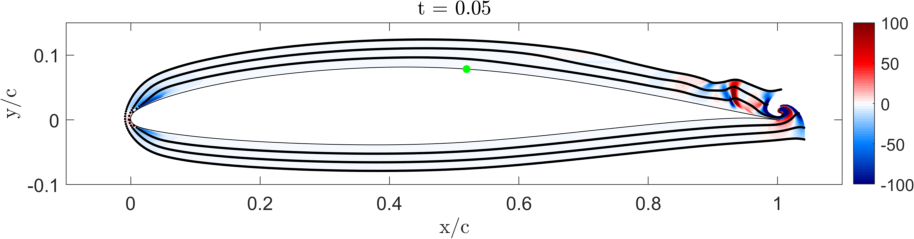}}
	\par
	\subfigure{\includegraphics[width=0.8\textwidth]{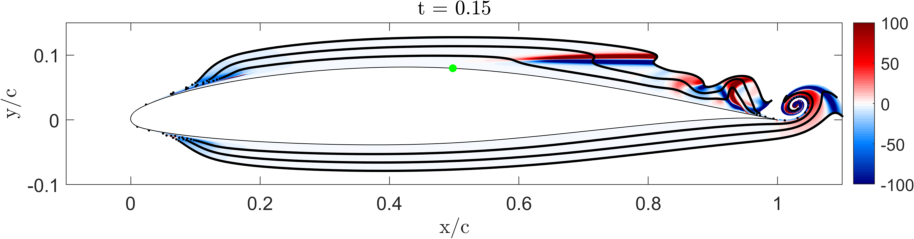}}
	\par
	\subfigure{\includegraphics[width=0.8\textwidth]{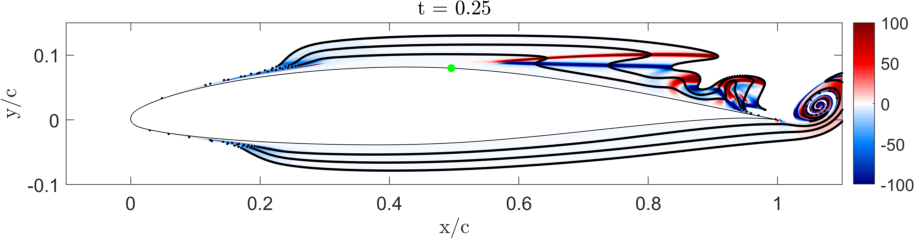}}
	\par
	\subfigure{\includegraphics[width=0.8\textwidth]{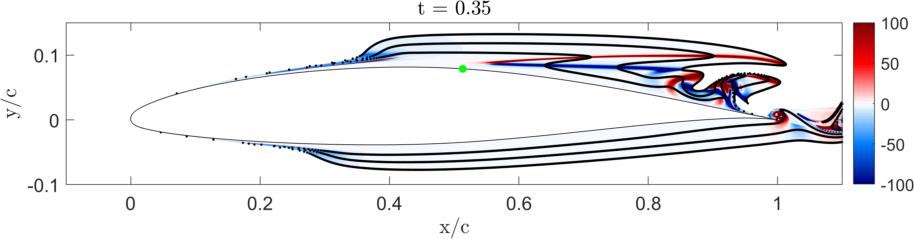}}
	\caption{Advection of material lines and the curvature field $\bar{\kappa}_{t_0}^{t_0+T}$ around the airfoil for different integration times. Zero-skin-friction point in green.}
	\label{fig:matline_airf}
\end{figure}

\begin{figure}[htp]
	\subfigure{\includegraphics[height=0.32\textwidth]{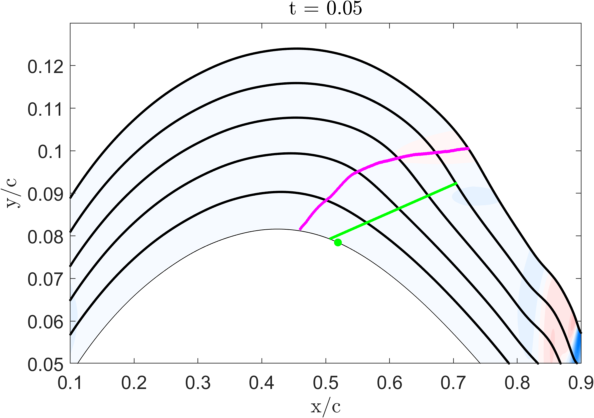}}
	\hfill
	\subfigure{\includegraphics[height=0.32\textwidth]{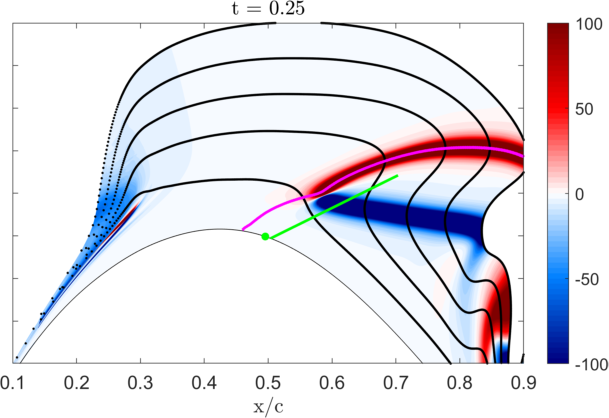}}
	\caption{Advected curvature scalar field with material lines (black) and the Lagrangian backbone of separation $\mathcal{B}(t)$ (magenta). Zero-skin-friction point and linear separation line in green. Y-axis stretched.}
	\label{fig:kappa_airf}
\end{figure}

The advected curvature change field at the final time, together with a set of
material lines, is shown in Figure \ref{fig:matline_airf} for different
integration times. Multiple spikes emerge on the suction side of the airfoil: a
dominant ridge evolves along the separating shear layer and several smaller
spikes appear within the separated recirculation region.  On the pressure
(bottom) side, the flow remains attached and the fluid tracers are advected
without breaking away from the boundary until the trailing edge is reached.
Given that global separation occurs only on the suction side of the airfoil, we
focus our analysis on the upper section of the profile. A magnified view of the
curvature scalar field introduced in Figure \ref{fig:matline_airf}, together
with the backbone of separation $\mathcal{B}(t)$ and the asymptotic separation
profile is given in Figure \ref{fig:kappa_airf}. Note that the y-axis is
stretched to aid visibility of subtle features.

The backbone emerging at mid-cord is based on the upwelling of separating
material lines in the vicinity of the asymptotic separation line and intersects
the no-slip wall at $s_p/c$ = 0.46. This location is slightly upstream of the
asymptotic separation point at $x/c$ = 0.5. 
Additional curvature ridges are
detected within the separated recirculation region, but, given that
the boundary layer has already separated, are of little
interest for determining the start of Lagrangian flow separation.
%Additional curvature ridges are
%detected within the separated recirculation region, but the
%curvature  along these lines is weak and the lines disappear with time. They are therefore of little
%interest for determining the start of Lagrangian flow separation.

%\FloatBarrier

\subsubsection*{Spiking phenomenon and FTLE}

\begin{figure}[htp]
	\subfigure{\includegraphics[width=0.49\textwidth]{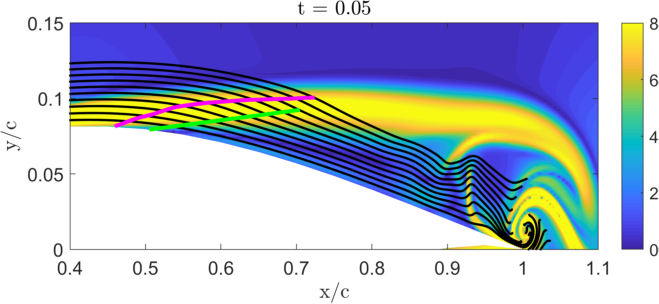}}
	\hfill
	\subfigure{\includegraphics[width=0.49\textwidth]{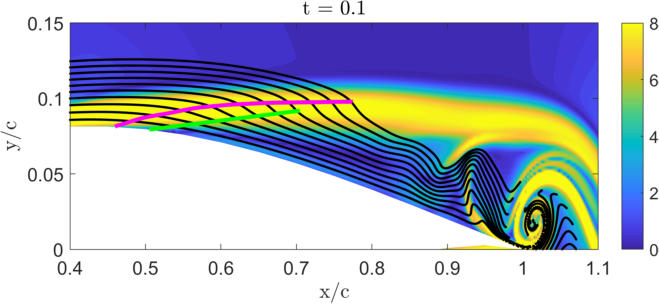}}
	\vskip\baselineskip	
	\subfigure{\includegraphics[width=0.49\textwidth]{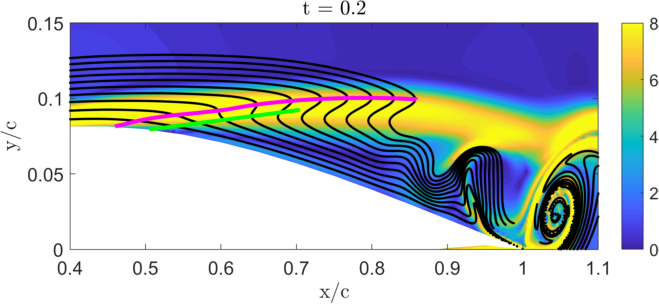}}
	\hfill
	\subfigure{\includegraphics[width=0.49\textwidth]{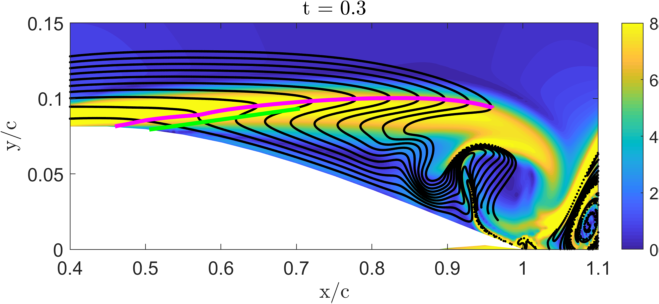}}
	\vskip\baselineskip
	\subfigure{\includegraphics[width=0.49\textwidth]{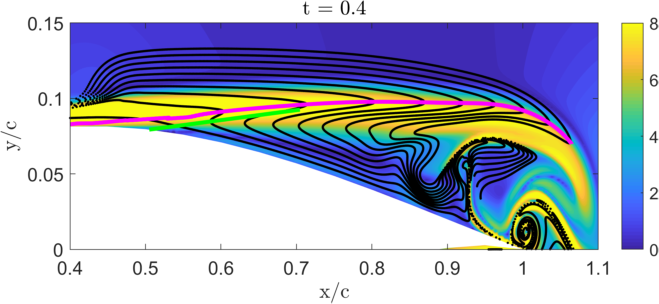}}
	\hfill
	\subfigure{\includegraphics[width=0.49\textwidth]{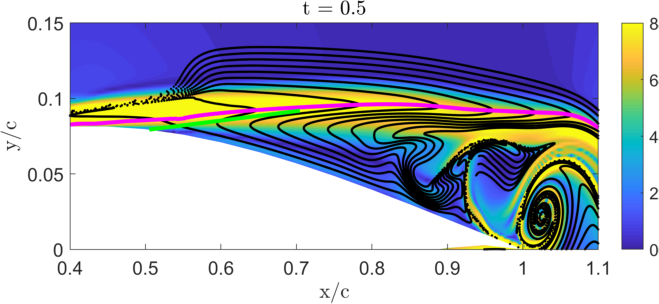}}
	\caption{Backward-time FTLE field (contour plot) computed from \textit{t} to \textit{t} $-$ \textit{T} over \textit{T} = 0.36. Advected material lines from 0 to \textit{t} in black and the Lagrangian backbone of separation in magenta. Asymptotic separation profile in green. Y-axis stretched.}
	\label{fig:ftle_airf}
\end{figure}

The relation between the Lagrangian backbone of separation, material lines, and the backward-time FTLE field is illustrated in Figure \ref{fig:ftle_airf} at different snapshots in time. 
To determine the  backward-time FTLE field,
again, we use an integration time interval equal to one
vortex shedding period (\textit{T} = 0.36). Material lines (black) 
and backbones (magenta) are advected forward in time from \textit{t} = 0 to \textit{t} = 0.05, 0.1,
0.2, 0.3, 0.4, and 0.5.

Similar to our findings for the cylinder flow, the material spike starts from the no-slip wall, crossing the FTLE ridge at short time scales. As the integration time increases, the material spike, along with the backbone of separation, aligns to the attracting FTLE ridge, which again governs the off-wall dynamics of the separated fluid tracers.
The separation picture is then completed by the asymptotic separation line (green) and its connection to backward-time FTLE ridge (see Figure \ref{fig:ftle_airf}). 

\subsubsection*{Extraction of Spiking Points}
\begin{figure}[htp]
	\subfigure{\includegraphics[width=0.3\textwidth]{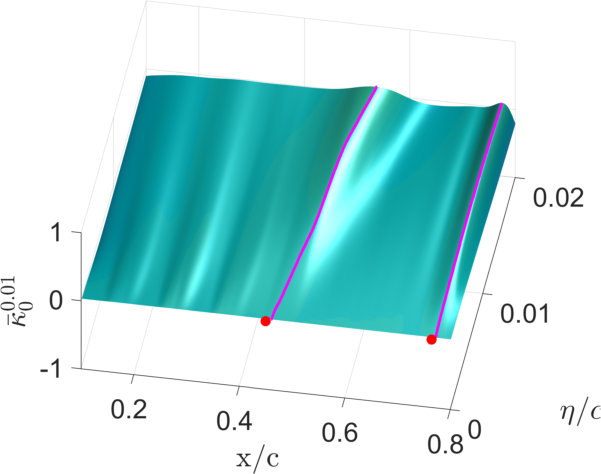}}
	\hfill
	\subfigure{\includegraphics[width=0.3\textwidth]{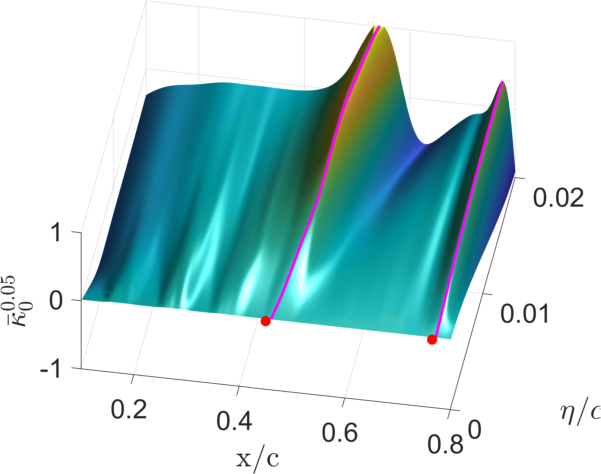}}
	\hfill
	\subfigure{\includegraphics[width=0.3\textwidth]{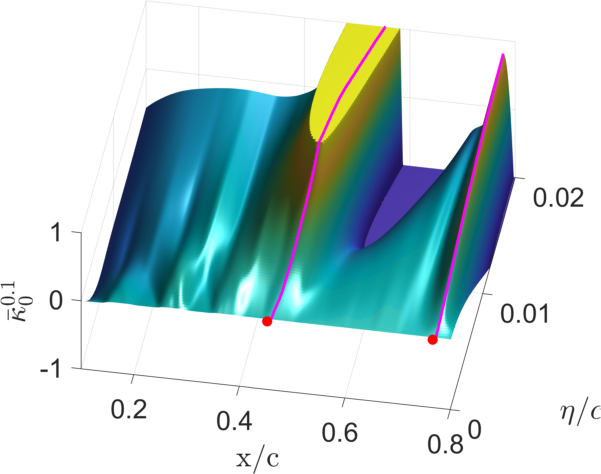}}
	\caption{Surface plot of the curvature scalar field $\bar{\kappa}_{t_0}^{t_0+T}$ for \textit{T} = 0.01, 0.05, and 0.10. Wall-normal coordinate on y-axis. Backbone in magenta, spiking point in red.}
	\label{fig:surf_normal_airf}
\end{figure}

\begin{figure}[htp]
	\centering
	\subfigure[]{\includegraphics[width=0.45\textwidth]{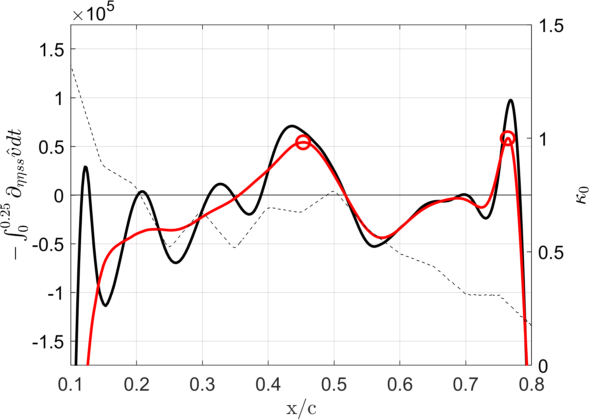}}
	\hspace{.5in}
	\subfigure[]{\includegraphics[width=0.42\textwidth]{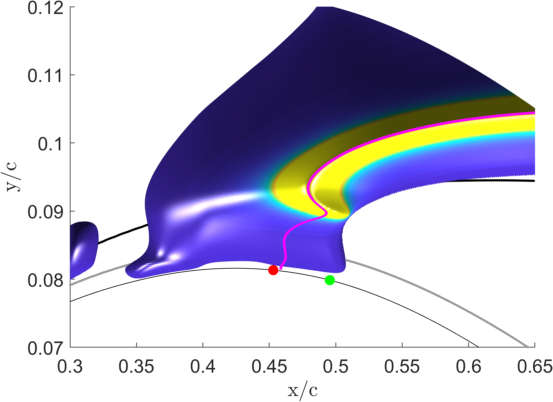}}
	\caption{(a) $-\int_{0}^{0.25}\partial_{\eta\eta ss}\hat{v} dt$ in black. Filtered results and Spiking points in red. Surface curvature $\kappa_0$ as dashed line. (b) Curvature change field $log(\bar{\kappa}_{0}^{0.25})$ at $\mathbf{x}_0$ with backbone (magenta) and Lagrangian spiking point (red) identified from Eulerian on-wall quantities. Displacement thickness in black and momentum thickness in grey. Zero-skin-friction point in green.}
	\label{fig:vnn_airf}
\end{figure}

We extract the spiking points of the separating airfoil flow from both their Lagrangian and wall-based Eulerian definitions. 
The curvature change field $\bar{\kappa}_{t_0}^{t_0+T}$ is given in Figure \ref{fig:surf_normal_airf} for three different integration intervals in $x$ and $\eta$ coordinates, where $\eta$ is the wall-normal distance. Besides the large ridges at mid-cord and at $x/c \approx 0.75$, a weak waviness in the curvature field exists upstream of the asymptotic separation point ($x/c$ = 0.5).
This oscillatory pattern is recovered in the Eulerian wall derivative $\int_{0}^{0.25}\partial_{\eta\eta ss}\hat{v} dt$, shown on left of Figure \ref{fig:vnn_airf}. 

According to the conditions specified in Table \ref{tab:sepPointLagrFormula},
spiking points are located at local maxima of the function
$-\int_{t_0}^{t_0+T}\partial_{\eta\eta ss}\hat{v} dt$, which identifies three
locations upstream of the separation point in Figure \ref{fig:vnn_airf} (a).
Weak curvature ridges are present at these locations
that we found are not contributing to material spiking and
fluid break away in the context of flow separation.  The oscillatory curvature
field and associated ridges correlate directly with the piece-wise linear curvature $\kappa_0$ of the
airfoil surface representation (dashed line) that is inherent
to the cubic spline boundary  representation used for the design of the airfoil.
The  three ridges are hence a geometric artifact
and should not be interpreted as significant spiking events.

We can reduce the oscillatory trend from the spline parametrization by
filtering the function $\int_{t_0}^{t_0+T}\partial_{\eta\eta ss}\hat{v} dt$
with a kernel width based on the approximate distance between two spline
segments. The filtered solution (red line) successfully recovers the underlying
correct function and identifies a single spiking point at $x/c$ = 0.45 (red
circle) upstream of the separation point.  

Figure \ref{fig:vnn_airf} (b) shows the curvature scalar field $\bar{\kappa}_{0}^{0.25}$ at $t_0$ in \textit{x} and \textit{y} coordinates with the ridge highlighted in magenta and the spiking point from Eulerian on-wall quantities in red. With the close match of the backbone-wall intersection at $x/c$ = 0.46 and the Eulerian criterion at $x/c$ = 0.45, we demonstrate that the spiking point can be extracted from on-wall based quantities even with approximate parametrization of the boundary, as used in engineering applications.

%\textcolor{magenta}{Bjoern: I would like to understand this and the following paragraph and the figure too} To visualize the subtleties of local fluid material upwelling over the curved airfoil surface, we trace a single material line from \textit{t} = 0 to \textit{t} = 0.5 and map the advected particles to the wall. Although this transformation to a curved coordinate system is incompatible with the methodology by Serra \textit{et al.} \cite{serra18}, it provides an \textit{ad hoc} strategy to illustrate the spiking phenomenon. 
%Figure \ref{fig:eta} shows that the line of fluid tracers initially inherits the oscillations from the velocity derivative $\partial_{\eta\eta}\hat{v}$ (see Figure \ref{fig:vnn_airf} (a)), but looses this signature for longer integration times and the particles are drawn towards the principal separation event at mid-cord. The main material spike along the global separation line is therefore not affected by local small-scale oscillations that are inherited from the piece-wise linear curvature of the spline.
%\begin{figure}[htp]
%	\centering
%	\includegraphics[width=0.50\textwidth]{img/Airfoil/eta.png}
%	\caption{Wall-normal distance of a material line for different integration times.}
%	\label{fig:eta}
%\end{figure}

\bigskip
%\subsubsection*{Treatment of the Leading Edge}

In Figure \ref{fig:matline_airf} and \ref{fig:kappa_airf}, we show that the
global separation of fluid particles traces back to the formation of an initial
material spike at mid-cord, shortly upstream of the asymptotic separation
point. For larger integration time intervals, however, an additional curvature
ridge with an origin at the leading edge is detected (see Figure
\ref{fig:kappa_airf}). This sharp spike in the material line occurs only in the
vicinity of the wall and has no direct connection to the global separation
event at mid-cord. Of course, when the leading edge ridge advects
to the location of the asymptotic separation manifold, the particles do 
follow this manifold. 

\begin{figure}[htp]
	\subfigure{\includegraphics[width=0.45\textwidth]{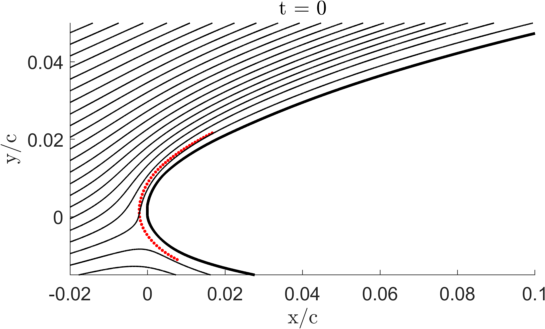}}
	\hfill
	\subfigure{\includegraphics[width=0.45\textwidth]{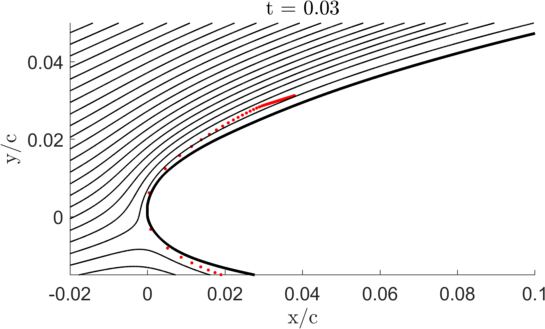}}
	\vskip\baselineskip
	\subfigure{\includegraphics[width=0.45\textwidth]{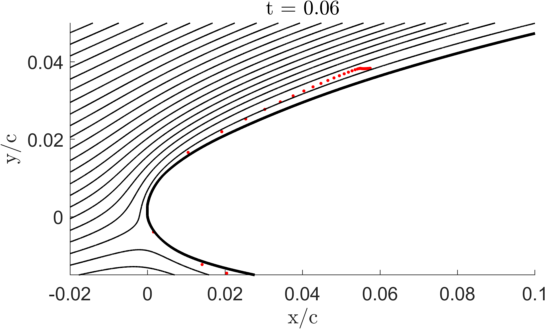}}
	\hfill	
	\subfigure{\includegraphics[width=0.45\textwidth]{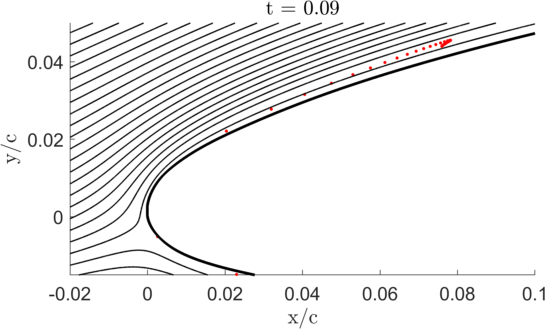}}
	\caption{Advection of a set of particles at the leading edge. Streamlines shown in black.}
	\label{fig:nose}
\end{figure}

Figure \ref{fig:nose} shows the advection of a Lagrangian particle sheet at the
leading edge to visualize the Lagrangian flow behavior at the stagnation point
in more detail: the
local contraction of the streamlines forces the fluid to accelerate and the
particles closer to the leading edge are initialized on streamlines with higher
velocity, leading to upwelling and folding of the material line.  Consequently,
the spike is rather the result of the displacement of streamlines by the
growing boundary layer and the associated normal velocity, than an event
associated with separating flow. 
We conclude that the leading edge spike formation is an artifact from the stagnation point flow and is not an indication of flow separation in this case. A more detailed analysis of this phenomenon will be subject to future studies.

\section{Conclusion} \label{conclusion}
We have investigated kinematic aspects of flow separation in external
aerodynamics by extracting the initial motion of upwelling fluid material from
the wall and its relation to the long-term attracting manifolds in the flow
field. While the wall-bounded kinematics are governed by the formation of a
material spike upstream of the asymptotic separation point and ejection of
particles in direction of the separation line, we show that the off-wall
trajectories of the fluid tracers are driven by attracting ridges in the
finite-time Lyapunov exponents. We have therefore obtained the complete pathway
of Lagrangian flow separation - from the initial upwelling at the spiking point
over ejection of particles along the asymptotic separation profile to the
attracting LCSs.

For the flow around a circular cylinder and a cambered NACA 65(1)-412 airfoil,
we extract the footprint of initial material upwelling, the spiking points, by
evaluating the curvature of Lagrangian fluid tracers and by extracting
high-order on-wall derivatives of the normal velocity as  proposed in
\cite{serra18}. An exact match of the Lagrangian and Eulerian criterion for the
start of material line spiking verifies the Eulerian criterion
and associated the principal location of material upwelling 
for the first time in test two test cases, i.e. the cylinder flow and 
the flow over an airfoil. 
For the latter, we recover the spiking point by appropriately filtering the spurious oscillations
in the velocity derivative induced by the spline-based boundary parametrization
of the NACA profile and show that this method is robust to noise.

With the ability to compute the birth of separation instantaneously from Eulerian on-wall data, the Lagrangian pathway from the spiking point to the asymptotic separation profile can be used as input parameters for dynamic flow controllers.
In future work, we aim to report on control of Lagrangian separation. We
will also  report on the correlation between the Lagrangian behavior
the backbone and boundary layer scaling laws.

\section*{Acknowledgments}
We gratefully acknowledge funding by the Air Force Office of Scientific Research under FA9550-16-1-0392 of the Flow Control Program and from Solar Turbines. Provision of subroutines for \textit{gslib} from Paul Fischer of the University of Illinois is greatly appreciated. Mattia Serra would like to acknowledge support from the Schmidt Science Fellowship \href{https://schmidtsciencefellows.org/}{(https://schmidtsciencefellows.org/)}.

%\section*{Appendix}
%\input{content/appendix}

\bibliographystyle{unsrt}
\bibliography{bib}

\end{document}